\documentclass[a4paper,12pt]{article}
\usepackage{mathrsfs}
\usepackage{amsmath, amsfonts, amssymb, mathrsfs, txfonts}
\usepackage{graphicx, subfigure}
\usepackage{wasysym}
\usepackage{color}
\usepackage{bm}
\usepackage{cases}
\usepackage{tikz}
\usepackage{multirow}
\usepackage[all]{xy}
\usepackage{graphicx}
\usepackage{pifont}
\usepackage{graphicx}
\usepackage{mathrsfs,psfrag,eepic,epsfig}
\usepackage{arydshln}
\usepackage{rotating}
\usepackage{cite}

\newtheorem{theorem}{Theorem}

\makeatletter \@addtoreset{equation}{section}

\newtheorem{corollary}{Corollary}

\newtheorem{proposition}{Proposition}
\newtheorem{remark}{Remark}

\makeatletter
\newfont{\footsc}{cmcsc10 at 8truept}
\newfont{\footbf}{cmbx10 at 8truept}
\newfont{\footrm}{cmr10 at 10truept}
\title{\bf{A symmetry-based method for constructing nonlocally related partial differential equation systems} }
\author{George W. Bluman\footnote{Electronic mail: bluman@math.ubc.ca} and Zhengzheng Yang\footnote{Electronic mail: yangzz@math.ubc.ca}
\\
\textit{Department of Mathematics, University of British
Columbia},\\
\textit{Vancouver, British Columbia V6T 1Z2, Canada}}

\date{}
\begin{document}
\maketitle

\noindent {\bf Abstract:}   Nonlocally related partial differential equation (PDE)
systems are important in the
analysis of a given PDE system. In particular, they are useful for seeking nonlocal symmetries. It is known that each local
conservation law of a given PDE system systematically yields a
nonlocally related PDE system.
In this paper, a new and complementary method for constructing
nonlocally related PDE systems is introduced.  In particular, it is shown
that each point symmetry of a PDE system systematically yields
a nonlocally related PDE system.  Examples include
nonlinear
reaction-diffusion equations, nonlinear diffusion equations and nonlinear wave equations. The considered nonlinear reaction-diffusion equations
have no local conservation laws.
Previously unknown nonlocal symmetries are
exhibited through our new symmetry-based method for two examples of nonlinear wave equations.

\section{Introduction}
A symmetry of a partial differential equation (PDE) system is any transformation of its solution manifold into itself, i.e., a symmetry transforms any solution of the PDE system to a solution of the same system.
In particular, continuous symmetries of a PDE system are continuous deformations of  its solutions to solutions of the same PDE system.
Consequently, continuous symmetries of PDE systems are defined topologically and hence not restricted to just point or local symmetries. Thus, in principle, any nontrivial PDE system
has symmetries.  The problem is to find and use symmetries.  Practically, to find a symmetry of a PDE system systematically, one is essentially restricted to transformations acting locally on some finite-dimensional space, whose variables   are not restricted to just the independent and dependent variables of the PDE system.  From this point of view,
local symmetries, whose infinitesimals depend at most on a finite number of derivatives of the dependent variables of the given PDE system, constitute only a subset of the total set of symmetries of a PDE system. Otherwise, there exist nonlocal symmetries of a PDE system \cite{Bca1,Olver,Bk1,Cant}.  However, when one directly  applies Lie's algorithm to find nonlocal symmetries, the coefficients of the infinitesimal generators should essentially involve integrals of the dependent variables  and their derivatives. It is difficult to set up and obtain solutions of corresponding determining equations for such coefficients.

In \cite{Bluman3}, a systematic procedure was introduced to seek nonlocal
symmetries (potential symmetries) for a given PDE system through
potential systems that naturally arise from its conservation laws. A
related heuristic approach to find nonlocal symmetries, called
quasilocal symmetries, was presented in \cite{Akh_Ibra1987,Akh_Ibra1991}, where rich sets of
examples were exhibited, especially those involving the gas dynamics equations.

An equivalent nonlocally PDE
system can play an important role in the analysis
of a given PDE system. Each solution of such a nonlocally related PDE
system yields a solution of the given PDE
system and, conversely, each solution of the given PDE system yields
a solution of the nonlocally related PDE system. These corresponding
solutions are obtained through connection formulas.  More importantly,
the relationship between the solutions is not one-to-one. Hence, for
a given PDE system,
one could be more successful when applying a standard method of
analysis, especially a coordinate
independent method, to a nonlocally related PDE system. For instance,
through a nonlocally related PDE
system, one can systematically find nonlocal symmetries and nonlocal
conservation laws of a given PDE system. It turns out that such
nonlocal symmetries and nonlocal conservation laws can arise as local
symmetries and local conservation laws of  nonlocally related PDE systems. Thus
any method depending on local symmetry analysis is valid for  nonlocally
related PDE systems. When nonlocal
symmetries can be found for a given PDE system, it may be possible to use such symmetries systematically to generate further exact solutions from its known solutions, to construct invariant solutions, to find linearizations, or find additional nonlocally related PDE systems (see, e.g., \cite{Cheviakov_JMP2008,Tsyfraa2005,B2007,Bluman_Cheviakov_Ivanova_JMP2006}).

 A systematic
procedure for finding nonlocally related PDE systems is presented in
\cite{Bca1} and references therein. Here one
constructs a tree of nonlocally
related systems that consists of potential systems and subsystems.
The potential systems arise naturally from local conservation laws.
However, open problems remain: How can one further systematically
extend a tree of nonlocally related PDE systems for a given PDE
system and, of particular importance, if the given system has no
local conservation law and no known nonlocally related subsystem?

         In this paper, we present a new systematic method for
constructing nonlocally related PDE systems for a given PDE system
through one-parameter Lie groups of point transformations that leave its solution manifold invariant (\textit{point symmetries}). In particular, we  show that a nonlocally
related PDE system (\textit{inverse potential system}) arises naturally from each point symmetry of a PDE system.
As a consequence, one is able to further extend the conservation law-based method
for the construction of trees of
nonlocally related PDE systems. For a given PDE system, we
show that nonlocally related PDE systems arising from its point
symmetries can also yield nonlocal symmetries.

 This paper is organized as follows. In Sect. 2, we introduce
the new systematic method to construct nonlocally
related PDE systems. In Sect. 3, the new method is used to construct
nonlocally related PDE systems for nonlinear reaction-diffusion equations, nonlinear diffusion equations and nonlinear wave equations. In
Sect. 4, the point symmetries for the inverse potential systems constructed in Sect. 3 are shown to yield nonlocal
symmetries for the considered example equations. Finally, in Sect. 5, the new results in
this paper are summarized and open problems are posed.

In this work, we use the package GeM for Maple \cite{Gem} for symmetry and conservation law analysis.

\section{New method: nonlocally related PDE systems arising from point symmetries}

Consider a PDE system of order $l$ with two independent variables $(x,t)$ and $m$ dependent variables $u=(u^1,\ldots,u^m)$ given by
\begin{equation} \label{SymB_given PDE}
R^\sigma[u]=R^\sigma(x,t,u,\partial u,\partial^2u,\ldots,\partial^lu)=0,\quad \sigma=1,\ldots,s,
\end{equation}
where $\partial^iu$ denotes the $i$-th order partial derivative of $u$.

A systematic conservation law-based method for constructing nonlocally related PDE systems of the PDE system  (\ref{SymB_given PDE}) was presented in \cite{Bluman3}. Here, the starting point is to use a nontrivial local conservation law of the PDE system (\ref{SymB_given PDE}):
\begin{equation}\label{a2}
 D_t\Phi[u]+ D_x\Psi[u]=0.
\end{equation}
Based on the conservation law (\ref{a2}), one constructs a corresponding nonlocally related PDE system (potential system) of the PDE system  (\ref{SymB_given PDE}) given by
\begin{equation}\label{a3}
  \left.
   \begin{array}{l}
    v_x=\Phi[u],\\
    v_t=-\Psi[u],\\
    R^\sigma[u]=0,\quad \sigma=1,\ldots,s.
   \end{array}
  \right.
\end{equation}

In this paper, instead of using a conservation law as the starting point, we present a new systematic method to construct a nonlocally related PDE system through the use of an admitted point symmetry as the starting point.

Suppose the PDE system (\ref{SymB_given PDE}) has a point symmetry with infinitesimal generator $\displaystyle\textbf{X}=\xi(x,t,u)\frac{\partial}{\partial x}+\tau(x,t,u)\frac{\partial}{\partial t}+\sum_{i=1}^m\eta^i(x,t,u)\frac{\partial}{\partial u^i}$. By introducing canonical coordinates corresponding to $\textbf{X}$:
\begin{equation}\label{SymB_canonical coordinates}
\begin{aligned}
&X=X(x,t,u),\\
&T=T(x,t,u),\\
&U^i=U^i(x,t,u),\quad i=1,\ldots,m,
\end{aligned}
\end{equation}
satisfying
\begin{equation}\label{SymB_canonical coordinates eq}
\begin{aligned}
&\textbf{X}X=0,\\
&\textbf{X}T=0,\\
&\textbf{X}U^1=1,\\
&\textbf{X}U^i=0,\quad i=2,\ldots,m,
\end{aligned}
\end{equation}
one maps $\textbf{X}$ into the canonical form $\textbf{Y}=\frac{\partial}{\partial U^1}$ while the PDE system (\ref{SymB_given PDE}) is mapped to an invertibly equivalent PDE system in terms of the canonical coordinates $(X,T,U)$ with $U=(U^1,\ldots,U^m)$. Since an invertible transformation maps a symmetry of a PDE system to a symmetry of the transformed system, $\textbf{Y}$ is the infinitesimal generator of a point symmetry of the invertibly equivalent PDE system. Consequently, the invertibly equivalent PDE system is invariant under translations in $U^1$. It follows that the invertibly equivalent PDE system is of the form
\begin{equation}\label{SymB_given PDE2}
 \hat{R}^\sigma(X,T,\hat{U},\partial U,\ldots,\partial^lU)=0,\quad \sigma=1,\ldots,s,
\end{equation}
where $\hat{U}=(U^2,\ldots,U^m)$.

Introducing two new variables $\alpha$ and $\beta$ for the first partial derivatives of $U^1$, one obtains the equivalent \textit{intermediate system} 
\begin{equation}\label{SymB_given PDE locally related}
\begin{aligned}
&\alpha=U^1_T,\\
&\beta=U^1_X,\\
&\tilde{R}^\sigma(X,T,\hat{U},\alpha,\beta,\partial\hat{U},\ldots,\partial^{l-1}\alpha,\partial^{l-1}\beta,\partial^l\hat{U})=0,\quad \sigma=1,\ldots,s,
\end{aligned}
\end{equation}
where $\tilde{R}^\sigma(X,T,\hat{U},\alpha,\beta,\partial\hat{U},\ldots,\partial^{l-1}\alpha,\partial^{l-1}\beta,\partial^l\hat{U})=0$  is obtained from $\hat{R}^\sigma(X,T,\hat{U},$ $\partial U,\ldots,$ $\partial^lU)=0$  after making the appropriate substitutions.  By construction, the intermediate system 
(\ref{SymB_given PDE locally related}) is  locally related to the PDE system
(\ref{SymB_given PDE2}),  and hence  locally related to  the given  PDE system 
(\ref{SymB_given PDE}).

Excluding the dependent variable $U^1$ from the intermediate system 
(\ref{SymB_given PDE locally related}), one obtains the \textit{inverse potential system} 
\begin{equation}\label{SymB_given PDE nonlocally related}
\begin{aligned}
&\alpha_X=\beta_T,\\
&\tilde{R}^\sigma(X,T,\hat{U},\alpha,\beta,\partial\hat{U},\ldots,\partial^{l-1}\alpha,\partial^{l-1}\beta,\partial^l\hat{U})=0,\quad \sigma=1,\ldots,s.
\end{aligned}
\end{equation}

Since the inverse potential system 
(\ref{SymB_given PDE nonlocally related}) is obtained by excluding $U^1$ from the intermediate system 
(\ref{SymB_given PDE locally related}), and $U^1$ cannot be expressed as a local function of $X$, $T$ and the remaining dependent variables $(\hat{U}, \alpha,\beta)$, and their derivatives, it follows that the inverse potential system 
(\ref{SymB_given PDE nonlocally related}) is nonlocally related to the PDE system 
(\ref{SymB_given PDE locally related}). In particular, the intermediate system 
(\ref{SymB_given PDE locally related}) is a potential system of the inverse potential system 
(\ref{SymB_given PDE nonlocally related}). Here, if $(\alpha, \beta, U^2, \ldots, U^m)=(f(x,t), g(x,t),$ $h^2(x,t),\ldots,h^m(x,t))$ solves the inverse potential system 
(\ref{SymB_given PDE nonlocally related}), there exists a family of functions $U^1=h^1(x,t)+C$, where $C$ is an arbitrary constant, such that $(\alpha, \beta, U^1, \ldots, U^m)=(f(x,t),$  $g(x,t),$ $h^1(x,t)+C, h^2(x,t),\ldots,h^m(x,t))$ solves the intermediate system 
(\ref{SymB_given PDE locally related}). By projection, $(U^1,\ldots, U^m)=(h^1(x,t)+C,h^2(x,t),\ldots,h^m(x,t))$ is a solution of the PDE system 
(\ref{SymB_given PDE2}). Thus the correspondence between the solutions of the inverse potential system 
(\ref{SymB_given PDE nonlocally related}) and those of the PDE system 
(\ref{SymB_given PDE2}) is not one-to-one. It follows that the inverse potential  system 
(\ref{SymB_given PDE nonlocally related}) is nonlocally related to  the PDE system 
(\ref{SymB_given PDE2}), and hence nonlocally related to the given PDE system  
(\ref{SymB_given PDE}).


Based on the above discussion, we have proved the following theorem.
\begin{theorem}\label{Th_symmtry based}\rm
Any point symmetry of a   PDE system 
(\ref{SymB_given PDE}) yields a nonlocally related  inverse potential system  given by the PDE system
(\ref{SymB_given PDE nonlocally related}). 
\end{theorem}
\begin{corollary}\rm
Consider a scalar PDE 
given by
\begin{equation}
u_t=F(x,t,u_1,...,u_n), \label{b1}
\end{equation}
where $u_i\equiv\frac{\partial^iu}{\partial x^i}$. Let $\beta=u_x$. Then the scalar PDE
\begin{equation}
\beta_t=D_xF(x,t,\beta,...,\beta_{n-1}) \label{bss1}
\end{equation}
 is locally related to the inverse potential system obtained from the invariance of the scalar PDE (\ref{b1}) under translations in $u$.
\end{corollary}
\medskip
\textbf{Proof.} Introducing new variables $\alpha$ and $\beta$ for the first partial derivatives of $u$, one obtains the  intermediate  system 

\begin{equation}\label{b2}
  \left.
   \begin{array}{l}
    \alpha=u_t,\\
    \beta=u_x,\\
    \alpha=F(x,t,\beta,...\beta_{n-1}),
   \end{array}
  \right.
\end{equation}
locally related to the PDE (\ref{b1}).
Excluding the dependent variable $u$ from the intermediate system 
(\ref{b2}), one obtains the inverse potential system 
\begin{equation}\label{b3}
  \left.
   \begin{array}{l}
   \alpha_x=\beta_t,\\
   \alpha=F(x,t,\beta,...,\beta_{n-1}).
   \end{array}
  \right.
\end{equation}
From Theorem 1, the inverse potential system 
(\ref{b3}) is nonlocally related to the PDE (\ref{b1}). Furthermore, one can exclude the dependent variable $\alpha$ from the inverse potential system 
(\ref{b3}) to obtain the subsystem given by the scalar PDE (\ref{bss1}).

Since the excluded variable $\alpha$ can be expressed from the equations of the inverse potential system 
(\ref{b3}) in terms of $\beta$ and its derivatives, the scalar PDE 
(\ref{bss1}) is locally related to the inverse potential system 
(\ref{b3}). \hfill$\square$

\noindent\begin{remark} \rm
A similar relationship between the scalar PDEs (\ref{b1}) and (\ref{bss1}) appears in \cite{Zhdanov2006}.
\end{remark}

\begin{remark} \rm
\textit{Connection between the symmetry-based method and the conservation law-based method.}  The symmetry-based method to obtain a nonlocally related PDE system does not require the existence of a nontrivial local conservation law of a given PDE system. Thus the new method is complementary to the conservation law-based method for constructing nonlocally related PDE systems. In particular, for the conservation law-based method, the constructed nonlocally related PDE system is a potential system of the given PDE system. For the symmetry-based method,  the  directly constructed intermediate system is locally related to the given PDE system. In turn, the intermediate system is a potential system of the inverse potential system. The symmetry-based method involves the reverse direction of the conservation law-based method.
\end{remark}
\begin{remark} \rm
\textit{The situation for a PDE system with at least three independent variables.} The symmetry-based method can be adapted  to a PDE system which  has at least three  independent variables. For simplicity, consider a scalar PDE 
with $n\geq3$ independent variables $x=(x^1,\ldots,x^n)$  and one dependent variable $u$:
\begin{equation}
R(x,u,\partial u,\partial^2u,\ldots,\partial^lu)=0. \label{SymB_multi givenPDE1}
\end{equation}
The general case can be considered in a similar way.
Suppose the scalar PDE 
(\ref{SymB_multi givenPDE1}) has a point symmetry with the infinitesimal generator $\textbf{X}$. In terms of canonical coordinates given by
\begin{equation}\label{SymB_multi canonical coordinates}
\begin{aligned}
&X^i=X^i(x,u),\quad i=1,\ldots,n,\\
&U=U(x,t,u),
\end{aligned}
\end{equation}
where
\begin{equation}\label{SymB_multi canonical coordinates eq}
\begin{aligned}
&\textbf{X}X^i=0,\quad i=1,\ldots,n,\\
&\textbf{X}U=1,
\end{aligned}
\end{equation}
the infinitesimal generator $\textbf{X}$ maps into the canonical form $\textbf{Y}=\frac{\partial}{\partial U}$. In terms of $(X,U)$ coordinates with $X=(X^1,\ldots,X^n)$, the scalar PDE 
(\ref{SymB_multi givenPDE1}) becomes  an invertibly  related  PDE 
of the form
\begin{equation}\label{SymB_multi given PDE}
 \hat{R}(X,\partial U,\partial^2U,\ldots,\partial^lU)=0.
\end{equation}

Introducing the new variables $\alpha=(\alpha^1,\ldots,\alpha^n)$ for the first partial derivatives of $U$, one obtains the equivalent locally related intermediate system 
\begin{equation}\label{SymB_multi locally related PDE}
   \begin{aligned}
    &\alpha^i=U_{X^i},~~i=1,\ldots,n,\\
    &\tilde{R}(X,\alpha,\partial \alpha\ldots,\partial^{l-1}\alpha)=0,
   \end{aligned}
\end{equation}
where $\tilde{R}(X,\alpha,\partial \alpha,\ldots,\partial^{l-1}\alpha)=0$  is obtained from $\hat{R}(X,\partial U,\partial^2U,\ldots,\partial^lU)=0$  after making the appropriate substitutions.
Excluding $U$ from the intermediate system 
(\ref{SymB_multi locally related PDE}), one obtains the inverse potential system 
\begin{equation}\label{SymB_multi nonlocally related PDE}
   \begin{aligned}
    &\alpha^i_{X^j}-\alpha^j_{X^i}=0,~~i,j=1,\ldots,n,\\
    &\tilde{R}(X,\alpha,\partial \alpha,\ldots,\partial^{l-1}\alpha)=0.
   \end{aligned}
\end{equation}
The inverse potential system 
(\ref{SymB_multi nonlocally related PDE}) is  nonlocally related to the scalar PDE 
(\ref{SymB_multi given PDE}), and hence nonlocally related to the scalar PDE 
(\ref{SymB_multi givenPDE1}). Moreover, since the PDE system 
(\ref{SymB_multi nonlocally related PDE})  has curl-type conservation laws, it could possibly yield nonlocal symmetries of the scalar PDE 
(\ref{SymB_multi givenPDE1}) from local symmetries of the inverse potential system (\ref{SymB_multi nonlocally related PDE}) \cite{Bca1,cb2,cb3}.
\end{remark}
\section{Examples of inverse potential systems}
In the previous section we introduced a new systematic symmetry-based method for constructing nonlocally related PDE systems  (inverse potential systems) of a given PDE system.  In this section, we illustrate this method by several examples.
\subsection{Nonlinear reaction-diffusion equations}
Consider the class of nonlinear reaction-diffusion equations
\begin{equation}\label{SymB_eg_nheat}
 u_t-u_{xx}=Q(u),
\end{equation}
where the reaction term $Q(u)$ is an arbitrary constitutive function with $Q_{uu}\neq 0$.
One can show that a nonlinear reaction-diffusion equation (\ref{SymB_eg_nheat}) has no nontrivial local conservation laws for any such $Q(u)$. \textit{Thus it is impossible to construct nonlocally related PDE systems for a nonlinear reaction-diffusion equation (\ref{SymB_eg_nheat}) by the conservation law-based method.}

On the other hand,  a nonlinear reaction-diffusion equation (\ref{SymB_eg_nheat}) has point symmetries. Thus one can construct nonlocally related PDE systems for a nonlinear reaction-diffusion equation (\ref{SymB_eg_nheat}) through the symmetry-based method introduced in Sect. 2. The point symmetry classification of the class of nonlinear reaction-diffusion equations (\ref{SymB_eg_nheat}) is presented in Table \ref{tab1} \cite{Dorodnitsyn1982}, modulo its group of equivalence transformations
\begin{equation}\label{ETSC_egET}
\begin{aligned}
   &\bar{x}=a_1x+a_2,\\
   &\bar{t}=a_1^2t+a_3,\\
   &\bar{u}=a_4u+a_5,\\
   &\bar{Q}=\frac{a_4}{a_1^2}Q.
 \end{aligned}
\end{equation}
\begin{table}[htbp]
\caption {Point symmetry classification for the class of nonlinear reaction-diffusion equations (\ref{SymB_eg_nheat})} \label{tab1} \medskip
\centering
\begin{tabular}{|c|c|l|}
  \hline
  $Q(u)$ & $\#$ & admitted point symmetries \\\hline
  arbitrary & 2 & $\textbf{X}_1=\frac{\partial}{\partial x}$, $\textbf{X}_2=\frac{\partial}{\partial t}$  \\\hline
  $u^a~(a\neq0,1)$ & 3 & $\textbf{X}_1$, $\textbf{X}_2$, $\textbf{X}_3=u\frac{\partial}{\partial u}
  -(a-1)t\frac{\partial}{\partial t}-\frac{a-1}{2}x\frac{\partial}{\partial x}$  \\\hline
  $e^{u}$ & 3 & $\textbf{X}_1$, $\textbf{X}_2$, $\textbf{X}_4=\frac{\partial}{\partial u}
  -t\frac{\partial}{\partial t}-\frac{1}{2}x\frac{\partial}{\partial x}$ \\\hline
  $u\ln u$ & 4 & $\textbf{X}_1$, $\textbf{X}_2$, $\textbf{X}_5=ue^t\frac{\partial}{\partial u},
  \textbf{X}_6=2e^t\frac{\partial}{\partial x}-xue^t\frac{\partial}{\partial u}$ \\\hline
\end{tabular}
\end{table}
\medskip

\noindent\textbf{(I) The case when $Q(u)$ is arbitrary}\medskip

\noindent For arbitrary $Q(u)$, a nonlinear  reaction-diffusion  equation (\ref{SymB_eg_nheat}) has the exhibited two point symmetries: $\textbf{X}_1$ and $\textbf{X}_2$. Therefore, using the symmetry-based method one can use interchanges of  $x$ and $u$ and also $t$ and $u$  to construct two inverse potential systems for a nonlinear reaction-diffusion equation (\ref{SymB_eg_nheat}).\medskip

\noindent\textbf{(I-a) Inverse potential system arising from $\textbf{X}_1$}\medskip

\noindent After an interchange of the variables $x$ and $u$, a nonlinear reaction-diffusion equation (\ref{SymB_eg_nheat}) becomes the invertibly related PDE given by
\begin{equation}\label{SymB_eg_nheat xeq}
x_t=\frac{x_{uu}-Q(u)x_u^3}{x_u^2}.
\end{equation}

Corresponding to the invariance of PDE (\ref{SymB_eg_nheat xeq}) under translations of its dependent variable $x$, one introduces the variables $v$ and $w$ for the first partial derivatives of $x$ to obtain the locally related intermediate system
 \begin{equation}\label{SymB_eg_nheat xeq locally}
   \begin{aligned}
  &v=x_u,\\
  &w=x_t,\\
  &w=\frac{v_u-Q(u)v^3}{v^2}.
 \end{aligned}
\end{equation}
Excluding $x$ from the intermediate system (\ref{SymB_eg_nheat xeq locally}), one obtains the inverse potential system
 \begin{equation}\label{SymB_eg_nheat xeq nonlocally}
   \begin{aligned}
&v_t=w_u,\\
&w=\frac{v_u-Q(u)v^3}{v^2}.
 \end{aligned}
\end{equation}
Moreover, one can exclude $w$ from the inverse potential system (\ref{SymB_eg_nheat xeq nonlocally}) to obtain its locally related subsystem 
\begin{equation}\label{SymB_eg_nheat xeq nonlocally scalar}
v_t=\left(\frac{v_u-Q(u)v^3}{v^2}\right)_u.
\end{equation}
Since the scalar PDE (\ref{SymB_eg_nheat xeq nonlocally scalar}) is in a conservation law form and a nonlinear reaction-diffusion equation  (\ref{SymB_eg_nheat}) has no local conservation laws, it follows that there is no invertible transformation that relates  the scalar PDE (\ref{SymB_eg_nheat xeq nonlocally scalar}) and the nonlinear reaction-diffusion equation  (\ref{SymB_eg_nheat}). Consequently, the scalar PDE (\ref{SymB_eg_nheat xeq nonlocally scalar}) is nonlocally related to  the nonlinear reaction-diffusion equation (\ref{SymB_eg_nheat}).
\medskip

\noindent\textbf{(I-b) Inverse potential system arising from $\textbf{X}_2$}\medskip

\noindent After an interchange of  the variables  $t$ and $u$, a nonlinear reaction-diffusion equation (\ref{SymB_eg_nheat}) becomes
\begin{equation}\label{SymB_eg_nheat teq}
t_u^2-Q(u)t_u^3+t_u^2t_{xx}-2t_xt_ut_{xu}+t_x^2t_{uu}=0,
\end{equation}
which is not in solved form and has  mixed derivatives.

Corresponding to the invariance of PDE (\ref{SymB_eg_nheat teq}) under translations of its dependent variable $t$,  one introduces two new variables $\alpha=t_x$ and $\beta=t_u$ to obtain the locally related intermediate system 
 \begin{equation}\label{SymB_eg_nheat teq locally}
   \begin{aligned}
&\alpha=t_x,\\
&\beta=t_u,\\
&\beta^2-Q(u)\beta^3+\beta^2\alpha_x-2\alpha\beta\alpha_u+\alpha^2\beta_u=0.
 \end{aligned}
\end{equation}
Excluding $t$ from the intermediate system (\ref{SymB_eg_nheat teq locally}), one obtains a second inverse potential system for a nonlinear reaction-diffusion equation (\ref{SymB_eg_nheat}) given by
 \begin{equation}\label{SymB_eg nheat teq nonlocally}
   \begin{aligned}
   &\alpha_u-\beta_x=0,\\
   &\beta^2-Q(u)\beta^3+\beta^2\alpha_x-2\alpha\beta\alpha_u+\alpha^2\beta_u=0.
   \end{aligned}
\end{equation}

The  constructed inverse potential systems for a nonlinear reaction-diffusion equation (\ref{SymB_eg_nheat}) ($Q(u)$ is arbitrary) are illustrated in Figure \ref{fig1}.\medskip
\begin{figure}[htbp]\label{Fig_SymB_eg_nheat_arbitrary}
\begin{center}
\begin{tikzpicture}
\node (tex) at(3.5,0) {\boxed{(3.1)}}; 
\node (dvi) at(1.5,-1.5) {\boxed{(3.5)}}; 
\node (dvi1) at(5.5,-1.5) {\boxed{(3.9)}}; 
\draw[-] (tex)--(dvi); 
\draw[-] (tex)--(dvi1); 
\end{tikzpicture}
\caption{Constructed inverse potential systems for a nonlinear reaction-diffusion equation (\ref{SymB_eg_nheat}) ($Q(u)$ is arbitrary).} \label{fig1}
\end{center}
\end{figure}

\noindent\textbf{(II) Inverse potential system arising from $\textbf{X}_3$ when $Q(u)=u^3$}\medskip

\noindent When $Q(u)=u^a$, ($a\neq0,1$),  the corresponding class of nonlinear reaction-diffusion equations (\ref{SymB_eg_nheat}) has one additional point symmetry $\textbf{X}_3$. For simplicity, we consider the case when $a=3$, i.e., $Q(u)=u^3$. The general case is considered in a similar way. Canonical coordinates induced by $\textbf{X}_3$ are given by
 \begin{equation}\label{SymB_eg nheat u3 ccordi}
   \begin{aligned}
   &X=xu,\\
  & T=\frac{t}{x^2},\\
  & U=-\ln x.
   \end{aligned}
\end{equation}
In $(X,T,U)$ coordinates, the  corresponding nonlinear reaction-diffusion  equation (\ref{SymB_eg_nheat}) becomes the invertibly related PDE
 \begin{equation}\label{SymB_eg nheat u3 invereq}
   \begin{aligned}
   &-3U_X^2-2XU_X^3-X^3U_X^3-U_X^2U_T+10TU_X^2U_T+U_{XX}-4TU_TU_{XX}\\
   &+4T^2U_T^2U_{XX}+4T^2U_X^2U_{TT}+4TU_XU_{TX}-8T^2U_XU_TU_{TX}=0.
   \end{aligned}
\end{equation}
Accordingly, introducing the new variables $\alpha=U_X$ and $\beta=U_T$, one obtains the locally related intermediate system
 \begin{equation}\label{SymB_eg nheat u3 locally}
   \begin{aligned}
   &\alpha=U_X,\\
   &\beta=U_T,\\
  & -3\alpha^2-2X\alpha^3-X^3\alpha^3-\alpha^2\beta+10T\alpha^2\beta+\alpha_{X}-4T\beta\alpha_{X}\\
  &+4T^2\beta^2\alpha_{X}+4T^2\alpha^2\beta_{T}+4T\alpha\beta_{X}-8T^2\alpha\beta\beta_{X}=0.
   \end{aligned}
\end{equation}
Excluding $U$ from the intermediate system (\ref{SymB_eg nheat u3 locally}), one obtains  a third inverse potential system of the corresponding nonlinear reaction-diffusion equation (\ref{SymB_eg_nheat}) given by
 \begin{equation}\label{SymB_eg nheat u3 nonlocally}
   \begin{aligned}
  & \alpha_T=\beta_X,\\
   & -3\alpha^2-2X\alpha^3-X^3\alpha^3-\alpha^2\beta+10T\alpha^2\beta+\alpha_{X}-4T\beta\alpha_{X}\\
  &+4T^2\beta^2\alpha_{X}+4T^2\alpha^2\beta_{T}+4T\alpha\beta_{X}-8T^2\alpha\beta\beta_{X}=0.
   \end{aligned}
\end{equation}

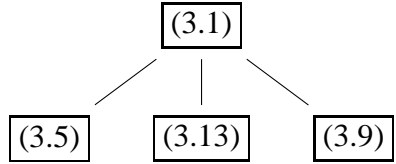
\begin{figure}[htbp]
\begin{center}
\begin{tikzpicture}
\node (tex) at(3.5,0) {\boxed{(3.1)}}; 
\node (tex1) at(3.5,-1.5) {\boxed{(3.13)}}; 
\node (dvi) at(1.5,-1.5) {\boxed{(3.5)}}; 
\node (dvi1) at(5.5,-1.5) {\boxed{(3.9)}}; 
\draw[-] (tex)--(tex1);
\draw[-] (tex)--(dvi); 
\draw[-] (tex)--(dvi1); 
\end{tikzpicture}
\caption{Constructed inverse potential systems for the nonlinear reaction-diffusion equation (\ref{SymB_eg_nheat}) ($Q(u)=u^3$).}\label{fig2}
\end{center}
\end{figure}

The constructed inverse potential systems for the nonlinear reaction-diffusion equation (\ref{SymB_eg_nheat}) ($Q(u)=u^3$) are illustrated in Figure \ref{fig2}.\medskip

Moreover, since the PDE systems (\ref{SymB_eg_nheat xeq nonlocally}) and (\ref{SymB_eg nheat u3 nonlocally})  do not have the same number of point symmetries, it follows that there is no invertible transformation relating these two PDE systems. Hence, the PDE systems (\ref{SymB_eg_nheat xeq nonlocally}) and (\ref{SymB_eg nheat u3 nonlocally}) are nonlocally related.
Similarly, the PDE systems (\ref{SymB_eg nheat teq nonlocally}) and (\ref{SymB_eg nheat u3 nonlocally}) are also nonlocally related.\medskip

\noindent\textbf{(III) Inverse potential system arising from $\textbf{X}_4$ when $Q(u)=e^u$}\medskip

\noindent When $Q(u)=e^u$, the nonlinear  reaction-diffusion equation (\ref{SymB_eg_nheat}) has one additional point symmetry $\textbf{X}_4$. Canonical coordinates induced by $\textbf{X}_4$ are given by
 \begin{equation}\label{SymB_eg nheat eu ccordi}
   \begin{aligned}
  & X=u+2\ln x,\\
   &T=\frac{t}{x^2},\\
   &U=-2\ln x.
   \end{aligned}
\end{equation}
In $(X,T,U)$ coordinates, the  corresponding nonlinear reaction-diffusion equation (\ref{SymB_eg_nheat}) becomes  the invertibly related PDE
 \begin{equation}\label{SymB_eg nheat eu inverteq}
   \begin{aligned}
 &  -2U_X^2-2U_X^3-e^XU_X^3-U_X^2U_T+6TU_X^2U_T+4U_{XX}-8TU_TU_{XX}\\
  & +4T^2U_T^2U_{XX}+4T^2U_X^2U_{TT}+8TU_XU_{TX}-8T^2U_XU_TU_{TX}=0.
   \end{aligned}
\end{equation}
It follows that the introduction of the new variables $\phi=U_X$ and $\psi=U_T$   yields the locally related intermediate system 
 \begin{equation}\label{SymB_eg nheat eu locally}
  \begin{aligned}
  & \phi=U_X,\\
  & \psi=U_T,\\
  & -2\phi^2-2\phi^3-e^X\phi^3-\phi^2\psi+6T\phi^2\psi+4\phi_{X}-8T\psi\phi_{X}\\
  &+4T^2\psi^2\phi_{X}+4T^2\phi^2\psi_{T}+8T\phi\psi_{X}-8T^2\phi\psi\psi_{X}=0.
   \end{aligned}
\end{equation}
Excluding $U$ from the intermediate system (\ref{SymB_eg nheat eu locally}), one obtains a third inverse potential system of the corresponding nonlinear reaction-diffusion (\ref{SymB_eg_nheat}) given by
 \begin{equation}\label{SymB_eg nheat eu nonlocally}
   \begin{aligned}
 &  \phi_T=\psi_X,\\
    & -2\phi^2-2\phi^3-e^X\phi^3-\phi^2\psi+6T\phi^2\psi+4\phi_{X}-8T\psi\phi_{X}\\
  &+4T^2\psi^2\phi_{X}+4T^2\phi^2\psi_{T}+8T\phi\psi_{X}-8T^2\phi\psi\psi_{X}=0.
   \end{aligned}
\end{equation}

The constructed inverse potential systems for the nonlinear reaction-diffusion equation (\ref{SymB_eg_nheat}) ($Q(u)=e^u$) are illustrated in Figure \ref{fig3}.\medskip
\begin{figure}[htbp]
\begin{center}
\begin{tikzpicture}
\node (tex) at(3.5,0) {\boxed{(3.1)}}; 
\node (dvi) at(1.5,-1.5) {\boxed{(3.5)}}; 
\node (dvi1) at(5.5,-1.5) {\boxed{(3.9)}}; 
\node (dvi2) at(3.5,-1.5) {\boxed{(3.17)}}; 
\draw[-] (tex)--(dvi); 
\draw[-] (tex)--(dvi1); 
\draw[-] (tex)--(dvi2); 
\end{tikzpicture}
\caption{Constructed inverse potential systems for the nonlinear reaction-diffusion equation (\ref{SymB_eg_nheat}) ($Q(u)=e^u$).} \label{fig3}
\end{center}
\end{figure}

Moreover,  since the PDE systems (\ref{SymB_eg_nheat xeq nonlocally}) and (\ref{SymB_eg nheat eu nonlocally})  do not have the same number of point symmetries, it follows that there is no invertible transformation relating these two PDE systems. Hence, the PDE systems (\ref{SymB_eg_nheat xeq nonlocally}) and (\ref{SymB_eg nheat eu nonlocally}) are nonlocally related.
Similarly, the PDE systems (\ref{SymB_eg nheat teq nonlocally}) and (\ref{SymB_eg nheat eu nonlocally}) are also nonlocally related.\medskip

\noindent\textbf{(IV) The case when $Q(u)=u\ln u$}\medskip

\noindent When $Q(u)=u\ln u$, the nonlinear  reaction-diffusion equation (\ref{SymB_eg_nheat}) has two additional point symmetries $\textbf{X}_5$ and $\textbf{X}_6$. \medskip

\noindent\textbf{(IV-a) Inverse potential system arising from $\textbf{X}_5$ }\medskip

\noindent Canonical coordinates induced by $\textbf{X}_5$ are given by
 \begin{equation}\label{SymB_eg nheat ulnu ccordina}
   \begin{aligned}
 &  X=x,\\
  & T=t,\\
   &U=e^{-t}\ln u.
   \end{aligned}
\end{equation}
In $(X,T,U)$ coordinates, the corresponding nonlinear reaction-diffusion  equation (\ref{SymB_eg_nheat}) becomes
 \begin{equation}\label{SymB_eg nheat ulnu inverteq}
   U_T=U_{XX}+e^{T}U_X^2.
\end{equation}
Introducing the new variables $p=U_X$ and $q=U_T$, one obtains the locally related intermediate system 
 \begin{equation}\label{SymB_eg nheat ulnu locally}
   \begin{aligned}
  & p=U_X,\\
   &q=U_T,\\
   &q=p_X+e^{T}p^2.
   \end{aligned}
\end{equation}
Excluding $U$ from the intermediate system (\ref{SymB_eg nheat ulnu locally}), one obtains the inverse potential system of the corresponding nonlinear reaction-diffusion (\ref{SymB_eg_nheat}) given by
 \begin{equation}\label{SymB_eg nheat ulnu nonlocally}
   \begin{aligned}
  & p_T=q_X,\\
   &q=p_X+e^{T}p^2.
   \end{aligned}
\end{equation}
Moreover, excluding $q$ from the inverse potential  system (\ref{SymB_eg nheat ulnu nonlocally}), one obtains the locally related subsystem of the inverse potential system (\ref{SymB_eg nheat ulnu nonlocally})  given by
 \begin{equation}\label{SymB_eg nheat ulnu nonlocally sub}
   p_T=p_{XX}+2e^{T}p p_X,
\end{equation}
which is in a conservation law form. Since the PDE (\ref{SymB_eg nheat ulnu nonlocally sub}) is in a conservation law form and any nonlinear reaction-diffusion equation  (\ref{SymB_eg_nheat}) has no local conservation laws, it follows that the  PDE (\ref{SymB_eg nheat ulnu nonlocally sub}) is nonlocally related to  the corresponding nonlinear reaction-diffusion equation (\ref{SymB_eg_nheat}).\medskip

\noindent\textbf{(IV-b) Inverse potential system arising from $\textbf{X}_6$}\medskip

\noindent Canonical coordinates induced by $\textbf{X}_6$ are given by
 \begin{equation}\label{SymB_eg nheat ulnu ccordi2}
   \begin{aligned}
 &  X=e^{\frac{x^2}{4}}u,\\
  & T=t,\\
   &U={\textstyle{1 \over 2}}e^{-t}x.
   \end{aligned}
\end{equation}
In $(X,T,U)$ coordinates, the corresponding nonlinear reaction-diffusion equation (\ref{SymB_eg_nheat}) becomes
 \begin{equation}\label{SymB_eg nheat ulnu inverteq2}
   U_T=\frac{e^{-2T}U_{XX}+2XU_X^3-4X\ln XU_X^3}{4U_X^2}.
\end{equation}
Introducing the new variables $r=U_X$ and $s=U_T$, one obtains the locally related intermediate system 
 \begin{equation}\label{SymB_eg nheat ulnu locally2}
   \begin{aligned}
 &  r=U_X,\\
  & s=U_T,\\
  &s=\frac{e^{-2T}r_{X}+2Xr^3-4X\ln Xr^3}{4r^2}.
   \end{aligned}
\end{equation}
Excluding $U$ from the intermediate system (\ref{SymB_eg nheat ulnu locally2}), one obtains the inverse potential system of the corresponding nonlinear reaction-diffusion (\ref{SymB_eg_nheat}) given by
 \begin{equation}\label{SymB_eg nheat ulnu nonlocally2}
   \begin{aligned}
 &  r_T=s_X,\\
  &s=\frac{e^{-2T}r_{X}+2Xr^3-4X\ln Xr^3}{4r^2}.
   \end{aligned}
\end{equation}
Excluding $s$ from the inverse potential system (\ref{SymB_eg nheat ulnu nonlocally2}), one obtains the locally related subsystem of the inverse potential system (\ref{SymB_eg nheat ulnu nonlocally2})  given by
 \begin{equation}\label{SymB_eg nheat ulnu nonlocally sub2}
   r_T=\left(\frac{e^{-2T}r_{X}+2Xr^3-4X\ln Xr^3}{4r^2}\right)_X.
\end{equation}
which is in a conservation law form. Since the PDE (\ref{SymB_eg nheat ulnu nonlocally sub2}) is in a conservation law form and any nonlinear reaction-diffusion equation  (\ref{SymB_eg_nheat}) has no local conservation laws, it follows that the  PDE (\ref{SymB_eg nheat ulnu nonlocally sub2}) is nonlocally related to  the corresponding nonlinear reaction-diffusion equation (\ref{SymB_eg_nheat}).\medskip

The constructed inverse potential systems for the nonlinear reaction-diffusion equation (\ref{SymB_eg_nheat}) ($Q(u)=u\ln u$)  are illustrated in Figure \ref{fig4}.
\begin{figure}[htbp]
\begin{center}
\begin{tikzpicture}
\node (tex) at(3.5,0) {\boxed{(3.1)}}; 
\node (dvi) at(2,-1.5) {\boxed{(3.5)}}; 
\node (dvi1) at(5,-1.5) {\boxed{(3.9)}}; 
\node (dvi3) at(-0.5,-1.5) {\boxed{(3.21)}}; 
\node (dvi4) at(7.5,-1.5) {\boxed{(3.26)}}; 
\draw[-] (tex)--(dvi); 
\draw[-] (tex)--(dvi1); 
\draw[-] (tex)--(dvi3); 
\draw[-] (tex)--(dvi4);
\end{tikzpicture}
\caption{Constructed inverse potential systems for the nonlinear reaction-diffusion equation (\ref{SymB_eg_nheat}) ($Q(u)=u\ln u$).} \label{fig4}
\end{center}
\end{figure}
\subsection{Nonlinear diffusion equations}
 As a second example, consider the class of scalar nonlinear diffusion equations
 \begin{equation}\label{SymB_eg nd given}
  v_t=K\left(v_x\right)v_{xx},
\end{equation}
where $K\left(v_x\right)$ is an arbitrary nonconstant constitutive function. The point symmetry classification of  its locally related class of PDE systems
 \begin{equation}\label{CLM_CLM_eg potential system1}
   \begin{aligned}
  &v_x=u,\\
  &v_t=K(u)u_x.
  \end{aligned}
\end{equation}
is listed in Table \ref{tab2} \cite{Bca1,Popovych_Ivanova_JPA2005a}, modulo its group of equivalence transformations  given by
\begin{equation}\label{NRS3_Eg_nonlocal symmetries ND equitrans}
\begin{aligned}
&\bar{t}=a_1t+a_2,\\
&\bar{x}=a_3x+a_4v+a_5,\\
&\bar{u}=\frac{a_6+a_7u}{a_3+a_4u},\\
&\bar{v}=a_6x+a_7v+a_8,\\
&\bar{K}=\frac{(a_3+a_4u)^2}{a_1}K,
\end{aligned}
\end{equation}
where $a_1$, $\ldots,$ $a_8$ are arbitrary constants with $a_1(a_3a_7-a_4a_6)\neq0$.

\begin{table}[ht]
\caption {Point symmetry classification for the class of PDE systems (\ref{CLM_CLM_eg potential system1})} \label{tab2} \medskip
\centering
\begin{tabular}{|c|c|l|}
  \hline
  $K(u)$ & $\#$ & admitted point symmetries \\\hline
  \multirow{2}{*}{arbitrary} & \multirow{2}{*}{4} &  $\textbf{Y}_1=\frac{\partial}{\partial x}$, $\textbf{Y}_2=\frac{\partial}{\partial t}$, $\textbf{Y}_3=x\frac{\partial}{\partial x}+
  2t\frac{\partial}{\partial t}+v\frac{\partial}{\partial v}$, \\
  & & $\textbf{Y}_4=\frac{\partial}{\partial v}$  \\\hline
  $u^\mu~(\mu\neq0)$ & 5 & $\textbf{Y}_1$, $\textbf{Y}_2$, $\textbf{Y}_3$, $\textbf{Y}_4$, $\textbf{Y}_5=x\frac{\partial}{\partial x}
  +\frac{2}{\mu}u\frac{\partial}{\partial u}+\left(1+\frac{2}{\mu}\right)v\frac{\partial}{\partial v}$  \\\hline
  $e^{u}$ & 5 & $\textbf{Y}_1$, $\textbf{Y}_2$, $\textbf{Y}_3$, $\textbf{Y}_4$, $\textbf{Y}_6=x\frac{\partial}{\partial x}
  +2\frac{\partial}{\partial u}+\left(2x+v\right)\frac{\partial}{\partial v}$ \\\hline
  \multirow{7}{*}{$u^{-2}$} & \multirow{7}{*}{$\infty$} & $\textbf{Y}_1$, $\textbf{Y}_2$, $\textbf{Y}_3$, $\textbf{Y}_4$, $\textbf{Y}_5$ $\left(\mu=-2\right),$\\
   & & $\textbf{Y}_7=-xv\frac{\partial}{\partial x}+(xu+v)u\frac{\partial}{\partial u}+2t\frac{\partial}{\partial v}$, \\
   & & $\textbf{Y}_8=-x(2t+v^2)\frac{\partial}{\partial x}+4t^2\frac{\partial}{\partial t}+u(6t+2xuv+v^2)\frac{\partial}{\partial u}$ \\
   & & $~~~~~~~~~+4tv\frac{\partial}{\partial v}$,\\
   & & $\textbf{Y}_\infty=F(v,t)\frac{\partial}{\partial x}-u^2G(v,t)\frac{\partial}{\partial u}$,  \\
   & & where $\left(F(v,t),G(v,t)\right)$ is an arbitrary solution \\
   & & of the linear system: $F_t=G_v$, $F_v=G$\\\hline
  \multirow{2}{*}{$\frac{1}{1+u^2}e^{\lambda \arctan u}$} &  \multirow{3}{*}{5} & $\textbf{Y}_1$, $\textbf{Y}_2$, $\textbf{Y}_3$, $\textbf{Y}_4$, \\
  & & $\textbf{Y}_{9}=v\frac{\partial}{\partial x}
 +\lambda t\frac{\partial}{\partial t}-\left(1+u^2\right)\frac{\partial}{\partial u}-x\frac{\partial}{\partial v}$ \\\hline
\end{tabular}
\end{table}
By projection of the symmetries in Table \ref{tab2}, one sees that for arbitrary $K\left(v_x\right)$, there are four point symmetries of a nonlinear diffusion equation (\ref{SymB_eg nd given}), namely, $\textbf{Y}_1=\frac{\partial}{\partial x}$, $\textbf{Y}_2=\frac{\partial}{\partial t}$, $\textbf{Y}_3=x\frac{\partial}{\partial x}+2t\frac{\partial}{\partial t}+v\frac{\partial}{\partial v}$ and $\textbf{Y}_4=\frac{\partial}{\partial v}$.\medskip

\noindent\textbf{(I) Inverse potential system arising from $\textbf{Y}_1$}\medskip

\noindent Since a nonlinear diffusion equation  (\ref{SymB_eg nd given}) is invariant under translations of its independent variable $x$, one can interchange $x$ and $v$ to generate an invertibly related PDE of a nonlinear diffusion equation (\ref{SymB_eg nd given}) given by
 \begin{equation}\label{SymB_eg nd given xeq}
  x_t=\frac{K\left(\frac{1}{x_v}\right)x_{vv}}{x_v^2}.
\end{equation}
Introducing new variables $w=x_v$ and $y=x_t$, one obtains the locally related intermediate system 
 \begin{equation}\label{SymB_eg nd given xeq locally}
   \begin{aligned}
&w=x_v,\\
&y=x_t,\\
&y=\frac{K\left(\frac{1}{w}\right)w_{v}}{w^2}.
 \end{aligned}
\end{equation}
Excluding $x$ from the intermediate system (\ref{SymB_eg nd given xeq locally}), one obtains the inverse potential system 
 \begin{equation}\label{SymB_eg nd given xeq nonlocally}
   \begin{aligned}
&w_t=y_v,\\
&y=\frac{K\left(\frac{1}{w}\right)w_{v}}{w^2}.
 \end{aligned}
\end{equation}

Moreover, one can exclude the variable $y$ from the inverse potential system (\ref{SymB_eg nd given xeq nonlocally}) to obtain the locally related subsystem of the inverse potential system (\ref{SymB_eg nd given xeq nonlocally}) given by
\begin{equation}\label{SymB_eg nd given xeq nonlocally pde}
  w_t=\left(\frac{K\left(\frac{1}{w}\right)w_{v}}{w^2}\right)_v.
\end{equation}

\noindent\textbf{(II) Inverse potential system arising from $\textbf{Y}_2$}\medskip

\noindent Since a nonlinear diffusion equation  (\ref{SymB_eg nd given}) is invariant under translations of its independent variable $t$, one can interchange
$t$ and $v$ to obtain an invertibly related PDE 
 given by
 \begin{equation}\label{SymB_eg nd given teq}
 t_v^2-K\left(-\frac{t_x}{t_v}\right)\left(2t_vt_xt_{xv}-t_x^2t_{vv}- t_v^2t_{xx}\right)=0.
\end{equation}
Introducing new variables $\alpha=t_v$ and $\beta=t_x$, one obtains the locally related intermediate system 
 \begin{equation}\label{SymB_eg nd given teq locally}
   \begin{aligned}
 &  \alpha=t_v,\\
  & \beta=t_x,\\
&\alpha^2-K\left(-\frac{\beta}{\alpha}\right)\left(2\alpha\beta\alpha_x-\beta^2\alpha_v-\alpha^2\beta_x\right)=0.
 \end{aligned}
\end{equation}
Excluding $t$ from the intermediate system (\ref{SymB_eg nd given teq locally}), one obtains the inverse potential system  
 \begin{equation}\label{SymB_eg nd given teq nonlocally}
   \begin{aligned}
&   \alpha_x=\beta_v,\\
&\alpha^2-K\left(-\frac{\beta}{\alpha}\right)\left(2\alpha\beta\alpha_x-\beta^2\alpha_v-\alpha^2\beta_x\right)=0.
 \end{aligned}
\end{equation}

\noindent\textbf{(III) Inverse potential system arising from $\textbf{Y}_3$}\medskip

\noindent Since a nonlinear diffusion equation  (\ref{SymB_eg nd given}) is invariant under the scaling symmetry generated by $\textbf{Y}_3=x\frac{\partial}{\partial x}+2t\frac{\partial}{\partial t}+v\frac{\partial}{\partial v}$, one can use a corresponding canonical coordinate transformation given by
\begin{equation}\label{SymB_eg nd given scale ccord}
   \begin{aligned}
&X=\frac{t}{x^2},\\
&T=\frac{v}{x},\\
&V=\ln x
 \end{aligned}
\end{equation}
to map a nonlinear diffusion equation (\ref{SymB_eg nd given}) into the invertibly related PDE
\begin{equation}\label{SymB_eg nd given scale inverteq}
   \begin{aligned}
&-V_XV_T^2+K\left(\frac{1+TV_T+2XV_X}{V_T}\right) \left(-4X V_T V_{TX}+V_{TT}+4 X V_XV_{TT} - V_T^2\right.\\
&-8 X^2 V_XV_T V_{TX} +4  X^2 V_X^2 V_{TT}\left.+2   X V_X V_T^2+4   X^2 V_T^2V_{XX}\right)=0.
\end{aligned}
\end{equation}
Introducing new variables $\phi=V_X$ and $\psi=V_T$, one obtains the locally related intermediate system 
\begin{equation}\label{SymB_eg nd given scale locally}
   \begin{aligned}
 &  \phi=V_X,\\
 &  \psi=V_T,\\
&-\phi \psi^2+K\left(\frac{1+T\psi+2X\phi}{\psi}\right)  \left(-4 X\psi \psi_{X} +  \psi_{T}+4  X \phi\psi_{T}  -  \psi^2 \right.\\
&\left.-8   X^2 \phi \psi\psi_{X} +4   X^2 \phi^2 \psi_{T}+2   X \phi \psi^2+4   X^2 \psi^2\phi_{X}\right)=0.
\end{aligned}
\end{equation}
Excluding $V$ from the intermediate system (\ref{SymB_eg nd given scale locally}), one obtains the inverse potential system 
 \begin{equation}\label{SymB_eg nd given scale nonlocally}
   \begin{aligned}
 &  \phi_T=\psi_X,\\
&-\phi \psi^2+K\left(\frac{1+T\psi+2X\phi}{\psi}\right)  \left(-4 X\psi \psi_{X} +  \psi_{T}+4  X \phi\psi_{T}  -  \psi^2 \right.\\
&\left.-8   X^2 \phi \psi\psi_{X} +4   X^2 \phi^2 \psi_{T}+2   X \phi \psi^2+4   X^2 \psi^2\phi_{X}\right)=0.
 \end{aligned}
\end{equation}

\noindent\textbf{(IV) Inverse potential system arising from $\textbf{Y}_4$}\medskip

\noindent From its invariance under translations of its  dependent variable $v$, one can apply directly the symmetry-based method to a nonlinear diffusion equation (\ref{SymB_eg nd given}). Letting  $u=v_x$, $z=v_t$, one obtains the corresponding locally related intermediate system 
 \begin{equation}\label{SymB_eg nd given locally}
   \begin{aligned}
  & u=v_x,\\
  & z=v_t,\\
  & z=K(u)u_x.
   \end{aligned}
\end{equation}
Excluding $v$ from the intermediate system (\ref{SymB_eg nd given locally}), one obtains  the inverse potential system 
 \begin{equation}\label{SymB_eg nd given nonlocally}
   \begin{aligned}
   &u_t=z_x,\\
  & z=K(u)u_x.
   \end{aligned}
\end{equation}
 Excluding $z$ from the inverse potential system (\ref{SymB_eg nd given nonlocally}), one obtains  the locally related subsystem of the inverse potential system (\ref{SymB_eg nd given nonlocally})  given by the class of nonlinear diffusion equations
 \begin{equation}\label{SymB_eg nd ueq}
  u_t=\left(K\left(u\right)u_{x}\right)_x.
\end{equation}

\noindent\textbf{(IV) Inverse potential system for a nonlinear diffusion equation (\ref{SymB_eg nd ueq})}\medskip

\noindent Now take a nonlinear diffusion equation (\ref{SymB_eg nd ueq}) as the given PDE. The point symmetry classification for the class of nonlinear diffusion equations (\ref{SymB_eg nd ueq}) is presented in Table \ref{tab3}  \cite{Ovsiannikov1959}, modulo its group of equivalence transformations given by
 \begin{equation}\label{SymB_add_equiv_trans_nd2}
   \begin{aligned}
   &\bar{t}=a_4t+a_1,\\
&\bar{x}=a_5x+a_2,\\
&\bar{u}=a_6u+a_3,\\
&\bar{K}=\frac{a_5^2}{a_4}K,
   \end{aligned}
\end{equation}
where $a_1$, $\ldots$, $a_6$ are arbitrary constants with $a_4a_5a_6\neq0$.

\begin{table}[ht]
\caption {Point symmetry classification for the class of nonlinear diffusion equations (\ref{SymB_eg nd ueq})} \label{tab3} \medskip
\centering
\begin{tabular}{|c|c|l|}
  \hline
   $K(u)$   & $\#$  &  admitted point symmetries \\\hline
  arbitrary & 3 & $\textbf{X}_1=\frac{\partial}{\partial x}$, $\textbf{X}_2=\frac{\partial}{\partial t}$, $\textbf{X}_3=x\frac{\partial}{\partial
x}+2t\frac{\partial}{\partial t}$  \\\hline
  $u^\mu ~(\mu\neq0)$    & 4 & $\textbf{X}_1$, $\textbf{X}_2$, $\textbf{X}_3$, $\textbf{X}_4=x\frac{\partial}{\partial
x}+\frac{2}{\mu}u\frac{\partial}{\partial u}$  \\\hline
  $e^{u}$   & 4 & $\textbf{X}_1$, $\textbf{X}_2$, $\textbf{X}_3$, $\textbf{X}_5=x\frac{\partial}{\partial x}+2\frac{\partial}{\partial
u}$ \\\hline
  $u^{-\frac{4}{3}}$ & 5 & $\textbf{X}_1$, $\textbf{X}_2$, $\textbf{X}_3$, $\textbf{X}_4$ $(\mu=-\frac{4}{3})$,  $\textbf{X}_6=x^2\frac{\partial}{\partial
x}-3xu\frac{\partial}{\partial u} $  \\
  \hline
\end{tabular}
\end{table}

There are three point symmetries of a nonlinear diffusion equation (\ref{SymB_eg nd ueq}) for arbitrary $K(u)$: $\textbf{X}_1=\frac{\partial}{\partial x}$, $\textbf{X}_2=\frac{\partial}{\partial t}$ and $\textbf{X}_3=x\frac{\partial}{\partial x}+2t\frac{\partial}{\partial t}$. Therefore, one can construct three inverse potential systems for a nonlinear diffusion equation (\ref{SymB_eg nd ueq}) through the symmetry-based method. Take $\textbf{X}_1$ for example. From its invariance  under translations in $x$, one can employ the hodograph transformation interchanging $x$
and $u$ to obtain the invertibly related PDE 
\begin{equation}\label{SymB_eg nd ueq xeq}
  x_t=-\left(\frac{K(u)}{x_u}\right)_u.
\end{equation}
Accordingly, letting $p=x_u$ and $q=x_t$, one obtains the locally related intermediate system 
\begin{equation}\label{SymB_eg nd ueq xeq locally}
   \begin{aligned}
&p=x_u,\\
&q=x_t,\\
&q=-\left(\frac{K(u)}{p}\right)_u.
   \end{aligned}
\end{equation}
Excluding the variable $x$ from the intermediate system (\ref{SymB_eg nd ueq xeq locally}), one obtains the inverse potential system 
\begin{equation}\label{SymB_eg nd ueq xeq nonlocally}
   \begin{aligned}
&p_t=q_u,\\
&q=-\left(\frac{K(u)}{p}\right)_u.
   \end{aligned}
\end{equation}
Finally, after excluding the variable $q$ from the inverse potential system (\ref{SymB_eg nd ueq xeq nonlocally}), one obtains the locally related subsystem of the inverse potential system  (\ref{SymB_eg nd ueq xeq nonlocally}) given by
\begin{equation}\label{SymB_eg nd ueq xeq nonlocally pde}
  p_t=-\left(\frac{K(u)}{p}\right)_{uu}.
\end{equation}

\begin{figure}[htbp]
\begin{center}
\begin{tikzpicture}
\node (tex) at(3.5,0) {\boxed{(3.28)}}; 
\node (dvi) at(2,-1.5) {\boxed{(3.37)}}; 
\node (dvi3) at(-0.5,-1.5) {\boxed{(3.33)}}; 
\node (dvi4) at(7.5,-1.5) {\boxed{(3.43)}}; 
\node (dvi1) at(5,-1.5) {\boxed{(3.41)}}; 
\node (dvi2) at(7.5,-3) {\boxed{(3.48)}};
\draw[-] (tex)--(dvi3);
\draw[-] (tex)--(dvi4);
\draw[-] (tex)--(dvi); 
\draw[-] (tex)--(dvi1); 
\draw[-] (dvi4)--(dvi2);
\end{tikzpicture}
\caption{Constructed inverse potential systems for a nonlinear diffusion equation (\ref{SymB_eg nd given}) ($K(v_x)$ is arbitrary).}\label{fig5}
\end{center}
\end{figure}
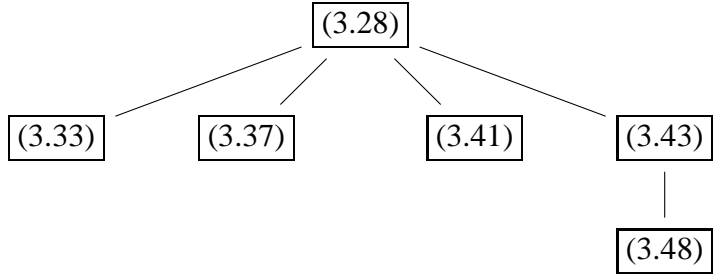

The constructed inverse potential systems for a nonlinear diffusion equation (\ref{SymB_eg nd given}) ($K(v_x)$ is arbitrary) are illustrated in Figure \ref{fig5}.
\subsection{Nonlinear wave equations}
As a third example, consider the class of nonlinear wave equations
\begin{equation}\label{SymB_eg nweq}
u_{tt}=(c^2(u)u_x)_x,
\end{equation}
with an arbitrary nonconstant constitutive function $c(u)$.

In \cite{B2007,Bca1}, it is shown that one can apply any invertible transformation to a PDE system with two or more dependent variables to seek additional nonlocally related subsystems of the given PDE system from exclusions of the resulting dependent variables. 
 Theorem 1 shows that the use of an invertible transformation that is a point symmetry of a given PDE system yields a nonlocally related PDE system (inverse potential system).  We now use point symmetries of the potential system of a nonlinear wave equation (\ref{SymB_eg nweq}) given by
\begin{equation}\label{SymB_eg nweq potential}
   \begin{aligned}
    &v_x=u_t,\\
    &v_t=c^2(u)u_x
   \end{aligned}
\end{equation}
to obtain additional nonlocally related PDE systems for a nonlinear wave equation (\ref{SymB_eg nweq}). For arbitrary $c(u)$, the potential system (\ref{SymB_eg nweq potential}) has the point symmetries  $\textbf{Y}_1=\frac{\partial}{\partial t}$, $\textbf{Y}_2=\frac{\partial}{\partial x}$, $\textbf{Y}_3=\frac{\partial}{\partial v}$, $\textbf{Y}_4=x\frac{\partial}{\partial x}+t\frac{\partial}{\partial t}$ and $\textbf{Y}_\infty$, where $\textbf{Y}_\infty$ represents the infinite number of point symmetries arising from the linearization of the potential system (\ref{SymB_eg nweq potential}) through the hodograph transformation (interchange of independent and dependent variables).

Due to its invariance under translations in $v$ and $t$, for arbitrary $c(u)$, the potential system (\ref{SymB_eg nweq potential}) has the point symmetry with the infinitesimal generator $\frac{\partial}{\partial v}-\frac{\partial}{\partial t}$.  Corresponding canonical coordinates yield the invertible point transformation
\begin{equation}\label{SymB_eg nweq poten tran}
\rho:~\left\{
   \begin{aligned}
  &  X=x,  \\
   & T=u,  \\
    &U=t+v,\\
    &V=v.
   \end{aligned}
   \right.
\end{equation}

The point transformation (\ref{SymB_eg nweq poten tran}) maps a potential system (\ref{SymB_eg nweq potential}) into the invertibly related PDE system
\begin{equation}\label{SymB_eg nweq poten invereq}
   \begin{aligned}
   & V_XU_T-V_TU_X-1=0,  \\
   & V_T+c^2(T)U_X-c^2(T)V_X=0,
   \end{aligned}
\end{equation}
which is invariant under translations in $U$ and $V$.

From the invariance of a PDE system (\ref{SymB_eg nweq poten invereq}) under translations in $V$, one introduces two new variables $A$ and $B$ for the first partial derivatives of $V$ to obtain the intermediate system 
\begin{equation}\label{SymB_eg nweq poten invereq_intermediate}
   \begin{aligned}
   & A=V_X,\\
   & B=V_T,\\
   & AU_T-BU_X-1=0,  \\
   & B+c^2(T)U_X-c^2(T)A=0,
   \end{aligned}
\end{equation}

Excluding $V$ from the intermediate system (\ref{SymB_eg nweq poten invereq_intermediate}), one obtains the inverse potential system
\begin{equation}\label{SymB_eg nweq poten invereq_inverse}
   \begin{aligned}
   & A_T=B_X,\\
   & AU_T-BU_X-1=0,  \\
   & B+c^2(T)U_X-c^2(T)A=0.
   \end{aligned}
\end{equation}

Since one can solve for $A$ and $B$ from the last two equations of the inverse potential system (\ref{SymB_eg nweq poten invereq_inverse}), it is straightforward to exclude $A$ and $B$ from the inverse potential system (\ref{SymB_eg nweq poten invereq_inverse}) to obtain its locally related scalar PDE
\begin{equation}\label{SymB_eg nweq Usub}
\begin{aligned}
&U_{TT}+c^2(T)\left(c^2(T)U_{XX}-U_{XX}U_T^2-U_{T
T}U_{X}^2-2U_{TX}+2U_{TX}U_{T}U_{X}\right)\\
&-2c(T)c'(T)\left(U_{X}-U_{X}^2U_{T}\right)=0.
 \end{aligned}
\end{equation}

In the next section, we prove that the PDE (\ref{SymB_eg nweq Usub}) is nonlocally related to the nonlinear wave equation (\ref{SymB_eg nweq}) through the symmetry classifications of these two classes of PDEs.
\begin{remark}\label{Rmk_SymB_eg nw1}\rm
Another equivalent straightforward method to obtain the scalar PDE (\ref{SymB_eg nweq Usub}) is by excluding $V$ directly from the PDE system (\ref{SymB_eg nweq poten invereq}) through cross-differentiation.
\end{remark}

The coordinates in the point transformation (\ref{SymB_eg nweq poten tran}) are  a choice of canonical coordinates corresponding to the point symmetry $\textbf{Y}_1$. The inverse potential system, arising from the invariance of the PDE system (\ref{SymB_eg nweq poten invereq}) under translations in $U$, yields the locally related scalar PDE
\begin{equation}\label{SymB_eg nweq Vsub vxu}
c(u)\left(v_x^2v_{uu}-2v_xv_uv_{ux}+v_{xx}v_u^2-c^2(u)v_{xx}\right)-2c'(u)v_x^2v_u=0.
\end{equation}
By interchanging $x$ and $v$, it is straightforward to show that the PDE (\ref{SymB_eg nweq Vsub vxu}) is invertibly related to the linear wave equation constructed in \cite{B2007}:
\begin{equation}\label{eg26}
x_{vv}=\left(c^{-2}(u)x_u\right)_u.
\end{equation}

\section{Examples of nonlocal symmetries arising from the symmetry-based method}

In the framework of nonlocally related PDE systems, nonlocal symmetries of a given PDE system (\ref{SymB_given PDE}) can arise from point symmetries of any PDE system in a  tree of nonlocally related PDE systems that includes (\ref{SymB_given PDE}).

In the conservation law-based method, from constructed nonlocally related PDE systems, three different types of nonlocal symmetries can be sought for a given PDE system  (\ref{SymB_given PDE}) \cite{Bca1}.
\begin{enumerate}
\renewcommand{\labelenumi}{\theenumi.}
\item Nonlocal symmetries arising from point symmetries of potential systems of (\ref{SymB_given PDE}).
\item Nonlocal symmetries arising from point symmetries of nonlocally related subsystems of (\ref{SymB_given PDE}).
\item Nonlocal symmetries arising from point symmetries of nonlocally related subsystems of potential systems of  (\ref{SymB_given PDE}).
\end{enumerate}

For Type 1, a point symmetry of  a potential system  of (\ref{SymB_given PDE}) yields a nonlocal symmetry of  (\ref{SymB_given PDE}) if and only if  the infinitesimal components corresponding to its given variables $(x,t,u)$ involve the nonlocal variables of the potential system. For Types 2 and 3, one must trace back to see whether the obtained point symmetry yields a nonlocal symmetry of  (\ref{SymB_given PDE}).

In the symmetry-based method, one can  seek further nonlocal symmetries arising from point symmetries of the constructed inverse potential systems as well as their subsystems.

In the previous section, we constructed several inverse potential systems for  nonlinear reaction-diffusion equations (\ref{SymB_eg_nheat}),  nonlinear diffusion equations (\ref{SymB_eg nd given}) and (\ref{SymB_eg nd ueq}),  and  nonlinear wave equations (\ref{SymB_eg nweq}).
For a nonlinear reaction-diffusion equation (\ref{SymB_eg_nheat}), one can show that each point symmetry of the constructed inverse potential systems yields no nonlocal symmetry of (\ref{SymB_eg_nheat}). In this section, it is shown that for  nonlinear diffusion equations (\ref{SymB_eg nd given}) and (\ref{SymB_eg nd ueq}), and  nonlinear wave equations (\ref{SymB_eg nweq}), nonlocal local symmetries do arise from some of the constructed inverse potential systems (or the locally related subsystems of such inverse potential systems). Most importantly, some previously unknown nonlocal symmetries are obtained for  the nonlinear wave equation (\ref{SymB_eg nweq}) when $c(u)=u^{-2}$ or $c(u)=u^{-\frac{2}{3}}$.
\subsection{Nonlocal symmetries of nonlinear diffusion equations}

In Tables \ref{tab2} and \ref{tab3}, we presented the point symmetry classifications for the classes of nonlinear diffusion equations (\ref{SymB_eg nd given}) and (\ref{SymB_eg nd ueq}).
\begin{proposition}\label{Prop_eg nd x6}\rm
 The symmetry $\textbf{X}_6$ yields a nonlocal symmetry of the corresponding nonlinear diffusion equation (\ref{SymB_eg nd given}) with $K(u)=u^{-\frac{4}{3}}$.
\end{proposition}
\textbf{Proof.} Suppose the symmetry $\textbf{X}_6$ yields a local symmetry of the nonlinear diffusion equation (\ref{SymB_eg nd given}) with $K(u)=u^{-\frac{4}{3}}$. Since the nonlinear diffusion equation (\ref{SymB_eg nd given}) and the potential system (\ref{CLM_CLM_eg potential system1}) are locally related,  $\textbf{X}_6$ must also yield a local symmetry $\hat{\textbf{X}}_6$ of the potential system (\ref{CLM_CLM_eg potential system1}). Consequently, there must exist a differential function $f[u,v]$ such that, in evolutionary form,  $\hat{\textbf{X}}_6=(-3xu-x^2u_x)\frac{\partial}{\partial u}+f[u,v]\frac{\partial}{\partial v}$ is a local symmetry of the potential system (\ref{CLM_CLM_eg potential system1}). Since $v_x=u$, $v_t=u^{-\frac{4}{3}}u_x$ and $u_t=(u^{-\frac{4}{3}}u_x)_x$,  one can restrict $f[u,v]$ to be of the form $f(x,t,u,v,u_x,u_{xx},...)$ depending on $x$, $t$, $u$ and the partial derivatives of $u$ with respect to $x$. Firstly, suppose $f[u,v]$  is of the form $f(x,t,u,v,u_x)$. Applying $\hat{\textbf{X}}_6^{(\infty)}$ to the potential system (\ref{CLM_CLM_eg potential system1}), one obtains
\begin{equation}\label{p51de}
  \left.
   \begin{array}{l}
   f_x+f_uu_x+f_vv_x+f_{u_x}u_{xx}=-3xu-x^2u_x,\\
   f_t+f_uu_t+f_vv_t+f_{u_x}u_{tx}=\frac{4}{3}(3xu+x^2u_x)u^{-\frac{7}{3}}u_x+D_x(-3xu-x^2u_x)u^{-\frac{4}{3}}
   \end{array}
  \right.
\end{equation}
on every solution of the potential system (\ref{CLM_CLM_eg potential system1}). After making appropriate substitutions and equating the coefficients of the term $u_{xx}$, one obtains $f_{u_x}=0$. By similar reasoning, one can show that $f(x,t,u,v,u_x,u_{xx},...)$ has no dependence on any partial derivative of $u$ with respect to $x$. Hence $f[u,v]$ is of the form $f(x,t,u,v)$. Consequently, if
$\textbf{X}_6$ yields a local symmetry of the nonlinear diffusion equation (\ref{SymB_eg nd given}) with $K(u)=u^{-\frac{4}{3}}$, then $\hat{\textbf{X}}_6$ must be a point symmetry of the corresponding potential system (\ref{CLM_CLM_eg potential system1}).

Comparing Tables \ref{tab2} and \ref{tab3}, one immediately sees that  symmetry $\textbf{X}_6$ does not yield a point symmetry of the corresponding potential system (\ref{CLM_CLM_eg potential system1}). This follows from the
fact that when
$K(u) = u^{-\frac{4}{3}}$, the potential system (\ref{CLM_CLM_eg potential system1}) has no point symmetry whose infinitesimal components corresponding to the variables $(x,t)$ are the same
as those for $\textbf{X}_6$.  Hence $\textbf{X}_6$ yields a nonlocal symmetry of the nonlinear diffusion equation (\ref{SymB_eg nd given}) with $K(u)=u^{-\frac{4}{3}}$.\hfill$\square$\medskip

Now consider the class of scalar PDEs (\ref{SymB_eg nd given xeq nonlocally pde}).
The equivalence transformations for this class arise from the six infinitesimal generators
\begin{equation}\label{SymB_eg nd given xeq nonlocally pde etran}
   \begin{aligned}
   &\textbf{E}_1=\frac{\partial}{\partial v},~~~~ \textbf{E}_2=\frac{\partial}{\partial w}+\frac{2K}{w}\frac{\partial}{\partial K},
   ~~~~\textbf{E}_3=w\frac{\partial}{\partial w}+2K\frac{\partial}{\partial K}, \\
  &\textbf{E}_4=v\frac{\partial}{\partial v}+2K\frac{\partial}{\partial K},~~~~\textbf{E}_5=t\frac{\partial}{\partial t}-K\frac{\partial}{\partial K},~~~~\textbf{E}_6=\frac{\partial}{\partial t}.
   \end{aligned}
\end{equation}
Thus the group of equivalence transformations for the class of PDEs  (\ref{SymB_eg nd given xeq nonlocally pde}) is given by
\begin{equation}\label{SymB_eg nd given xeq nonlocally pde etran2}
   \begin{aligned}
   &\bar{v}=a_3v+a_1,\\
   & \bar{t}=a_5t+a_6,\\
   &\bar{w}=a_4w+a_2,\\
  & \bar{K}=\frac{a_3^2(a_4w+a_2)^2}{a_5w^2}K,
   \end{aligned}
\end{equation}
where $a_1$, $\ldots,$ $a_6$ are arbitrary constants with $a_3a_4a_5\neq0$.

In Table \ref{tab4}, we present the point symmetry classification for the class of PDEs (\ref{SymB_eg nd given xeq nonlocally pde}),
modulo its group of equivalence transformations (\ref{SymB_eg nd given xeq nonlocally pde etran2}).\medskip

\begin{table}[ht]\label{Table_SymB_eg nd given xeq nonlocally pde}
\caption {Point symmetry classification for the class of PDEs (\ref{SymB_eg nd given xeq nonlocally pde})} \label{tab4} \medskip
\centering
\begin{tabular}{|c|c|c|l|l|}
  \hline
   \multirow{2}{*}{$K\left(1/w\right)$}  & \multirow{2}{*}{$K(u)$} & \multirow{2}{*}{$\#$}  & admitted point symmetries in & admitted point symmetries in\\
   & & & $(t,v,w)$ coordinates & $(t,v,u)$ coordinates\\\hline

   arbitrary  &arbitrary  & 3 & $\textbf{V}_1=\frac{\partial}{\partial t}$, $\textbf{V}_2=\frac{\partial}{\partial v}$, $\textbf{V}_3=2t\frac{\partial}{\partial t}+v\frac{\partial}{\partial v}$ &$\textbf{V}_1$, $\textbf{V}_2$, $\textbf{V}_3$\\\hline

  \multirow{2}{*}{$w^{-\mu}$}      &\multirow{2}{*}{$u^{\mu}$}    &  \multirow{2}{*}{4} & $\textbf{V}_1$, $\textbf{V}_2$, $\textbf{V}_3$, &$\textbf{V}_1$, $\textbf{V}_2$, $\textbf{V}_3$,\\
   & & &  $\textbf{V}_4=(2+\mu)v\frac{\partial}{\partial v}-2w\frac{\partial}{\partial w}$& $\textbf{V}_4=(2+\mu)v\frac{\partial}{\partial v}+2u\frac{\partial}{\partial u}$\\\hline

  \multirow{2}{*}{$w^{\frac{2}{3}}$}      &\multirow{2}{*}{$u^{-\frac{2}{3}}$}   & \multirow{2}{*}{5} & $\textbf{V}_1$, $\textbf{V}_2$, $\textbf{V}_3$, $\textbf{V}_4$ $(\mu=-\frac{2}{3})$,&$\textbf{V}_1$, $\textbf{V}_2$, $\textbf{V}_3$, $\textbf{V}_4$ $(\mu=-\frac{2}{3})$,\\
    & & & $\textbf{V}_5=3vw\frac{\partial}{\partial w}-v^2\frac{\partial}{\partial v}$& $\textbf{V}_5=-3uv\frac{\partial}{\partial u}-v^2\frac{\partial}{\partial v}$\\\hline

  \multirow{5}{*}{$w^{2}$}    &\multirow{5}{*}{$u^{-2}$} & \multirow{5}{*}{$\infty$} & $\textbf{V}_1$, $\textbf{V}_2$, $\textbf{V}_3$, $\textbf{V}_4$ $(\mu=-2)$, & $\textbf{V}_1$, $\textbf{V}_2$, $\textbf{V}_3$, $\textbf{V}_4$ $(\mu=-2)$, \\
   & & & $ \textbf{V}_6=-vw\frac{\partial}{\partial w}+2t\frac{\partial}{\partial v}$, & $\textbf{V}_6=uv\frac{\partial}{\partial u}+2t\frac{\partial}{\partial v}$,\\
    &  & &$ \textbf{V}_7=4t^2\frac{\partial}{\partial t}+4vt\frac{\partial}{\partial v}-(2t+v^2)w\frac{\partial}{\partial w}$,  & $ \textbf{V}_7=4t^2\frac{\partial}{\partial t}+4vt\frac{\partial}{\partial v}+(2t+v^2)u\frac{\partial}{\partial u}$,\\
    &   & &$ \textbf{V}_\infty=G(t,v)\frac{\partial}{\partial w}$, where $G(t,v)$ &$ \textbf{V}_\infty=-u^2G(t,v)\frac{\partial}{\partial u}$, where $G(t,v)$\\
     & & &   satisfies $G_t=G_{vv}$ & satisfies $G_t=G_{vv}$\\\hline

\multirow{1}{*}{$e^ww^2$}      &\multirow{1}{*}{$e^{\frac{1}{u}}u^{-2}$}    &  \multirow{1}{*}{4} & $\textbf{V}_1$, $\textbf{V}_2$, $\textbf{V}_3$, $\textbf{V}_8=v\frac{\partial}{\partial v}+2\frac{\partial}{\partial w}$ &$\textbf{V}_1$, $\textbf{V}_2$, $\textbf{V}_3$,  $\textbf{V}_8=v\frac{\partial}{\partial v}-2u^2\frac{\partial}{\partial u}$\\\hline
\end{tabular}
\end{table}

By similar reasoning as in the proof of Proposition \ref{Prop_eg nd x6}, one can show that, for $K(u)=u^{-\frac{2}{3}}$, the point symmetry  $\textbf{V}_5$
 of the PDE (\ref{SymB_eg nd given xeq nonlocally pde}) yields a nonlocal symmetry of the corresponding intermediate system (\ref{SymB_eg nd given xeq locally}), which is locally related to the  nonlinear diffusion equation (\ref{SymB_eg nd given}). Hence  $\textbf{V}_5$  yields a nonlocal symmetry of the nonlinear diffusion equation (\ref{SymB_eg nd given}) with $K(u)=u^{-\frac{2}{3}}$.

Moreover, comparing Tables \ref{tab3} and \ref{tab4},
one also sees that when $K(u) = u^{-\frac{2}{3}}$, since its infinitesimal component for the variable $u$ has an essential dependence on the variable $v$,
 the  symmetry $\textbf{V}_5$ of the corresponding PDE (\ref{SymB_eg nd given xeq nonlocally pde}) yields a nonlocal symmetry of the nonlinear diffusion equation (\ref{SymB_eg nd ueq}), which cannot be obtained through its potential system (\ref{CLM_CLM_eg potential system1}).
By similar reasoning, when $K(u) = u^{-2}$,  one can show that the symmetries
$\textbf{V}_6$ , $\textbf{V}_7$  and $\textbf{V}_\infty$  of the PDE (\ref{SymB_eg nd given xeq nonlocally pde}) yield nonlocal symmetries of the corresponding nonlinear diffusion equation (\ref{SymB_eg nd ueq}). In addition, when $K(u)=e^{\frac{1}{u}}u^{-2}$, one can show that $\textbf{V}_8$  yields a point symmetry $\tilde{\textbf{V}}_8=(x+2v)\frac{\partial}{\partial x}+v\frac{\partial}{\partial v}-2u^2\frac{\partial}{\partial u}$ of the potential system (\ref{CLM_CLM_eg potential system1}) whose infinitesimal component for the variable $x$ has an essential dependence on the variable $v$. Consequently, $\textbf{V}_8$ yields a nonlocal symmetry of the nonlinear diffusion equation (\ref{SymB_eg nd ueq}) when $K(u)=e^{\frac{1}{u}}u^{-2}$.


Next consider the class of PDEs (\ref{SymB_eg nd ueq xeq nonlocally pde}).
The equivalence transformations for this class  arise from the six infinitesimal generators
\begin{equation}\label{SymB_eg nd ueq xeq nonlocally pde etran}
   \begin{aligned}
  & \textbf{E}_1=\frac{\partial}{\partial u},~~~~ \textbf{E}_2=u\frac{\partial}{\partial u}+2K\frac{\partial}{\partial K},
   ~~~~\textbf{E}_3=p\frac{\partial}{\partial p}+2K\frac{\partial}{\partial K}, \\
   &\textbf{E}_4=t\frac{\partial}{\partial t}-K\frac{\partial}{\partial K},~~~~\textbf{E}_5=\frac{\partial}{\partial t},~~~~\textbf{E}_6=u^2\frac{\partial}{\partial u}-3up\frac{\partial}{\partial p}-2Ku\frac{\partial}{\partial K}.
   \end{aligned}
\end{equation}
Correspondingly, the six-parameter group of equivalence transformations for the PDE class (\ref{SymB_eg nd ueq xeq nonlocally pde}) 
is given by
\begin{equation}\label{SymB_eg nd ueq xeq nonlocally pde etran2}
   \begin{aligned}
   &\bar{u}=a_2u+a_1,\\
 &  \bar{t}=a_4t+a_5,\\
  & \bar{p}=a_3p,\\
   &\bar{K}=\frac{a_2^2a_3^2}{a_4}K,
   \end{aligned}
\end{equation}
and
\begin{equation}\label{SymB_eg nd ueq xeq nonlocally pde etran3}
   \begin{aligned}
   &\bar{u}=\frac{u}{1-a_6u},\\
   &\bar{t}=t,\\
  & \bar{p}=(1-a_6u)^3p,\\
   &\bar{K}=(1-a_6u)^2K,
   \end{aligned}
\end{equation}
where $a_1$, $\ldots,$ $a_6$ are arbitrary constants with $a_2a_3a_4\neq0$.

In Table \ref{tab5}, we present the point symmetry classification for the class of PDEs (\ref{SymB_eg nd ueq xeq nonlocally pde}),
modulo its group of equivalence  transformations given by (\ref{SymB_eg nd ueq xeq nonlocally pde etran2}) and (\ref{SymB_eg nd ueq xeq nonlocally pde etran3}).\medskip

\begin{table}[ht]
\caption {Point symmetry classification for the class of PDEs (\ref{SymB_eg nd ueq xeq nonlocally pde})} \label{tab5} \medskip
\centering
\begin{tabular}{|c|c|l|}
  \hline
   $K\left(u\right)$   & $\#$  & admitted point symmetries \\\hline
  arbitrary & 2 & $\textbf{W}_1=\frac{\partial}{\partial t}$, $\textbf{W}_2=2t\frac{\partial}{\partial t}+p\frac{\partial}{\partial p}$  \\\hline
  $u^{\mu}$    & 3 & $\textbf{W}_1$, $\textbf{W}_2$, $\textbf{W}_3=2u\frac{\partial}{\partial
u}+(\mu-2)p\frac{\partial}{\partial p}$  \\\hline
  $e^u$   & 3 & $\textbf{W}_1$, $\textbf{W}_2$, $\textbf{W}_4=2\frac{\partial}{\partial u}+p\frac{\partial}{\partial p}$ \\\hline
   \multirow{2}{*}{$\frac{1}{1+u^2}e^{\lambda \arctan u}$} & \multirow{2}{*}{3} & $\textbf{W}_1$, $\textbf{W}_2$,\\
 & & $\textbf{W}_5=2(1+u^2)\frac{\partial}{\partial u}-p(6u-\lambda)\frac{\partial}{\partial p}$  \\\hline
 $u^{-2}$    & 4 & $\textbf{W}_1$, $\textbf{W}_2$, $\textbf{W}_3$ $(\mu=-2)$, $\textbf{W}_{6}=u^2\frac{\partial}{\partial u}-3pu\frac{\partial}{\partial p}$  \\\hline
\end{tabular}
\end{table}

Similar to the situation in Proposition \ref{Prop_eg nd x6}, when $K(u)=\frac{1}{1+u^2}e^{\lambda \arctan u}$, the point symmetry $\textbf{W}_5$ of  the PDE (\ref{SymB_eg nd ueq xeq nonlocally pde}) yields a nonlocal symmetry of the corresponding intermediate system (\ref{SymB_eg nd ueq xeq locally}), which is locally related to  the nonlinear
diffusion  equation (\ref{SymB_eg nd ueq}). Hence $\textbf{W}_5$  yields a nonlocal symmetry of the nonlinear
diffusion  equation (\ref{SymB_eg nd ueq}) with $K(u)=\frac{1}{1+u^2}e^{\lambda \arctan u}$.  By similar reasoning, the symmetry $\textbf{W}_6$  also yields a nonlocal symmetry of  the nonlinear
diffusion  equation (\ref{SymB_eg nd ueq}) with $K(u)=u^{-2}$.

Taking the equivalence transformation (\ref{SymB_eg nd ueq xeq nonlocally pde etran3}) into consideration, one can obtain more nonlocal symmetries for the class of nonlinear
diffusion  equations (\ref{SymB_eg nd ueq}) from the corresponding class of PDEs (\ref{SymB_eg nd ueq xeq nonlocally pde}). In particular, the equivalence transformation (\ref{SymB_eg nd ueq xeq nonlocally pde etran3}) maps $u^\mu$ into $\bar{u}^\mu(1+a_6\bar{u})^{-(\mu+2)}$, $e^u$ into $(1+a_6\bar{u})^{-2}e^{\frac{\bar{u}}{1+a_6\bar{u}}}$. Moreover, the symmetries $\textbf{W}_3$ and $\textbf{W}_4$ are mapped into $\bar{\textbf{W}}_3$ and $\bar{\textbf{W}}_4$ respectively. One can show that when $K(u)=u^\mu(1+a_6u)^{-(\mu+2)}$, $\bar{\textbf{W}}_3=2u(1+a_6u)\frac{\partial}{\partial u}-p(6a_6u-\mu+2)\frac{\partial}{\partial p}$; when $K(u)=(1+a_6u)^{-2}e^{\frac{u}{1+a_6u}}$, $\bar{\textbf{W}}_4=2(1+a_6u)^2\frac{\partial}{\partial u}-p(6a_6^2u+6a_6-1)\frac{\partial}{\partial p}$. Similar to the situation in Proposition \ref{Prop_eg nd x6}, one can show that $\bar{\textbf{W}}_3$ and $\bar{\textbf{W}}_4$ yield nonlocal symmetries of the corresponding nonlinear
diffusion  equations (\ref{SymB_eg nd ueq}).

\begin{remark}\rm
Comparing Tables \ref{tab2} and \ref{tab5}, one concludes that when $K(u)=\frac{1}{1+u^2}e^{\lambda \arctan u}$, the nonlocal symmetry yielded by $\textbf{W}_5$  corresponds to the nonlocal symmetry yielded by $\textbf{Y}_9$.  When $K(u)=u^{-2}$, the nonlocal symmetry yielded by  $\textbf{W}_6$ corresponds to a nonlocal symmetry yielded by $\textbf{Y}_{\infty}$.
\end{remark}
\subsection{Nonlocal symmetries of  nonlinear wave equations}
We now use the subsystem (\ref{SymB_eg nweq Usub}), locally related to the inverse potential system (\ref{SymB_eg nweq poten invereq_inverse}), to obtain previously unknown nonlocal symmetries for the class of nonlinear wave equations  (\ref{SymB_eg nweq}).

In \cite{Ames_Lohner_Adams_JNM1981},  the point symmetry classification was obtained for the class of nonlinear wave equations  (\ref{SymB_eg nweq}), which is presented in Table \ref{tab6},
modulo its group of equivalence transformations
\begin{equation}\label{SymB_eg nweq etran}
  \begin{aligned}
   &\bar{x}=a_1x+a_4,\\
 &  \bar{t}=a_2t+a_5,\\
  & \bar{u}=a_3u+a_6,\\
   &\bar{c}=\frac{a_1}{a_2}c,
   \end{aligned}
\end{equation}
where $a_1$, $\ldots,$ $a_6$ are arbitrary constants with $a_1a_2a_3\neq0$.\medskip
\begin{table}[ht]
\caption {Point symmetry classification for the class of nonlinear wave equations (\ref{SymB_eg nweq})} \label{tab6} \medskip
\centering
\begin{tabular}{|c|c|l|}
  \hline
   $c(u)$& $\#$ & admitted point symmetries \\\hline
  arbitrary & 3 & $\textbf{X}_1=\frac{\partial}{\partial x}$, $\textbf{X}_2=\frac{\partial}{\partial t}$, $\textbf{X}_3=x\frac{\partial}{\partial x}+t\frac{\partial}{\partial
t}$  \\\hline
  $u^\mu$ & 4 & $\textbf{X}_1$, $\textbf{X}_2$, $\textbf{X}_3$, $\textbf{X}_4=\mu x\frac{\partial}{\partial
x}+u\frac{\partial}{\partial u}$  \\\hline
  $e^{u}$ & 4 & $\textbf{X}_1$, $\textbf{X}_2$, $\textbf{X}_3$, $\textbf{X}_5=x\frac{\partial}{\partial x}+\frac{\partial}{\partial
u}$ \\\hline
  $u^{-2}$ & $5$ & $\textbf{X}_1$, $\textbf{X}_2$, $\textbf{X}_3$, $\textbf{X}_4$ $(\mu=-2)$, $\textbf{X}_6=t^2\frac{\partial}{\partial
t}+tu\frac{\partial}{\partial u}$  \\\hline
  $u^{-\frac{2}{3}}$ & $5$ & $\textbf{X}_1$, $\textbf{X}_2$, $\textbf{X}_3$, $\textbf{X}_4$ $(\mu=-\frac{2}{3})$, $\textbf{X}_7=x^2\frac{\partial}{\partial
x}-3xu\frac{\partial}{\partial u}$  \\
  \hline
\end{tabular}
\end{table}

The equivalence transformations  for the PDE class (\ref{SymB_eg nweq Usub}) arise from the five infinitesimal generators
\begin{equation}\label{SymB_eg nweq Usub etran1}
   \begin{aligned}
   &\textbf{E}_1=\frac{\partial}{\partial T},~~~~ \textbf{E}_2=\frac{\partial}{\partial X},
   ~~~~\textbf{E}_3=\frac{\partial}{\partial U}, \\
   &\textbf{E}_4=T\frac{\partial}{\partial T}+X\frac{\partial}{\partial X}+U\frac{\partial}{\partial U},~~~~\textbf{E}_5=-T\frac{\partial}{\partial T}+X\frac{\partial}{\partial X}+c\frac{\partial}{\partial c}.
   \end{aligned}
\end{equation}
Correspondingly, the five-parameter group of equivalence transformations for the class of PDEs (\ref{SymB_eg nweq Usub}) is given by
\begin{equation}\label{SymB_eg nweq Usub etran2}
   \begin{aligned}
 &  \bar{T}=\frac{a_4}{a_5}T+a_1,\\
  & \bar{X}=a_4a_5X+a_2,\\
   &\bar{U}=a_4U+a_3,\\
   &\bar{c}=a_5c,
   \end{aligned}
\end{equation}
where $a_1$, $\ldots,$ $a_5$ are arbitrary constants with $a_4a_5\neq0$.

The point symmetry classification for the class of PDEs (\ref{SymB_eg nweq Usub}), modulo its equivalence
transformations (\ref{SymB_eg nweq Usub etran2}),  is presented in Table \ref{tab7}.\medskip

\begin{table}[ht]
\caption {Point symmetry classification for the class of PDEs (\ref{SymB_eg nweq Usub})} \label{tab7} \medskip
\centering
\begin{tabular}{|c|c|c|l|l|}
  \hline
   \multirow{2}{*}{$c(T)$}& \multirow{2}{*}{$c(u)$}& \multirow{2}{*}{$\#$}  & admitted point symmetries in & $(x,u)$ components of admitted\\
   & & & $(X,T,U)$ coordinates &  symmetries\\\hline

  \multirow{2}{*}{arbitrary} & \multirow{2}{*}{arbitrary} & \multirow{2}{*}{3} & $\textbf{W}_1=\frac{\partial}{\partial U}$, $\textbf{W}_2=\frac{\partial}{\partial X}$, &$\check{\textbf{W}}_2=\frac{\partial}{\partial x}$,\\
  & &   & $\textbf{W}_3=(X+\int^Tc^2(\xi)d\xi)\frac{\partial}{\partial X}+U\frac{\partial}{\partial U}$ & $\check{\textbf{W}}_3=(x+\int^uc^2(\xi)d\xi)\frac{\partial}{\partial x}$  \\\hline

  \multirow{3}{*}{$T^\mu$} &\multirow{3}{*}{$u^\mu$} & \multirow{2}{*}{4} & $\textbf{W}_1$, $\textbf{W}_2$, $\textbf{W}_3$ $(c(T)=T^\mu)$,  &  $\check{\textbf{W}}_2$, $\check{\textbf{W}}_3$ $(c(u)=u^\mu)$,\\
   & &   &$\textbf{W}_4=T\frac{\partial}{\partial T}+(2\mu+1)X\frac{\partial}{\partial X}$ &$\check{\textbf{W}}_4=u\frac{\partial}{\partial u}+(2\mu+1)x\frac{\partial}{\partial x}$\\
   & & & $~~~~~~~~+(\mu+1)U\frac{\partial}{\partial U}$&\\\hline

  \multirow{2}{*}{$e^{T}$} &\multirow{2}{*}{$e^{u}$} & \multirow{2}{*}{4} & $\textbf{W}_1$, $\textbf{W}_2$, $\textbf{W}_3$ $(c(T)=e^{T})$,& $\check{\textbf{W}}_2$, $\check{\textbf{W}}_3$ $(c(u)=e^{u})$,\\
   & &   &$\textbf{W}_5=\frac{\partial}{\partial T}+2X\frac{\partial}{\partial X}+U\frac{\partial}{\partial U}$ &$\check{\textbf{W}}_5=\frac{\partial}{\partial u}+2x\frac{\partial}{\partial x}$\\\hline

  \multirow{3}{*}{$T^{-2}$} & \multirow{3}{*}{$u^{-2}$} &\multirow{3}{*}{5} & $\textbf{W}_1$, $\textbf{W}_2$, &$\check{\textbf{W}}_1$, $\check{\textbf{W}}_2$,  \\
  & & & $\textbf{W}_3$ $(c(T)=T^{-2})$, $\textbf{W}_4$ $(\mu=-2)$,  &$\check{\textbf{W}}_3$ $(c(u)=u^{-2})$, $\check{\textbf{W}}_4$ $(\mu=-2)$,\\
  & &   &$\textbf{W}_6=U^2\frac{\partial}{\partial U}+TU\frac{\partial}{\partial T}-\frac{U}{T^3}\frac{\partial}{\partial X}$ &$\check{\textbf{W}}_6=u(t+v)\frac{\partial}{\partial u}-\frac{t+v}{u^3}\frac{\partial}{\partial x}$ \\ \hline

  \multirow{4}{*}{$T^{-\frac{2}{3}}$} & \multirow{4}{*}{$u^{-\frac{2}{3}}$} &\multirow{4}{*}{5} & $\textbf{W}_1$, $\textbf{W}_2$,  &$\check{\textbf{W}}_1$, $\check{\textbf{W}}_2$, \\
  & &   &$\textbf{W}_3$ $(c(T)=T^{-\frac{2}{3}})$, $\textbf{W}_4$ $(\mu=-\frac{2}{3})$,&$\check{\textbf{W}}_3$ $(c(u)=u^{-\frac{2}{3}})$, $\check{\textbf{W}}_4$ $(\mu=-\frac{2}{3})$,\\
  & &   &$ \textbf{W}_7=(XT-3T^{\frac{2}{3}})\frac{\partial}{\partial T}$ & $\check{\textbf{W}}_7=(xu-3u^{\frac{2}{3}})\frac{\partial}{\partial u}$\\
  & &   & $~~~~~~~~+(XT^{-\frac{1}{3}}-\frac{X^2}{3})\frac{\partial}{\partial X}$&  $~~~~~~~~+(xu^{-\frac{1}{3}}-\frac{x^2}{3})\frac{\partial}{\partial x}$\\\hline
\end{tabular}
\end{table}

\begin{remark}\label{Rmk_symb nw V change}\rm
In order to determine whether a symmetry $\textbf{W}$ of a PDE (\ref{SymB_eg nweq Usub}) yields a nonlocal symmetry of the corresponding nonlinear wave equation (\ref{SymB_eg nweq}), we need to trace back to the nonlinear wave equation (\ref{SymB_eg nweq}) using the PDE system (\ref{SymB_eg nweq poten invereq}). Since the PDE (\ref{SymB_eg nweq Usub}) excludes the dependent variable $V$ of the PDE system (\ref{SymB_eg nweq poten invereq}), we need to investigate how the variable $V$ changes under the action induced by $\textbf{W}$.
Since $\rho^{-1}(\frac{\partial}{\partial V})=\frac{\partial}{\partial v}-\frac{\partial}{\partial t}$, where $\rho^{-1}$ is the inverse of the transformation (\ref{SymB_eg nweq poten tran}), the infinitesimal components for the variables $x$ and $u$ remain invariant when tracing back. This is why we only present the $(x,u)$ components of admitted symmetries in  Table \ref{tab7}.
\end{remark}

\begin{proposition}\label{Prop_symb nw nonlocal symm}\rm
The symmetries $\textbf{W}_6$ and $\textbf{W}_7$ yield nonlocal symmetries of the corresponding potential systems (\ref{SymB_eg nweq potential}).
\end{proposition}
\textbf{Proof.} If the symmetry $\textbf{W}_6$ yields a local symmetry $\hat{\textbf{W}}_6$ of the  potential system (\ref{SymB_eg nweq potential}) with $c(u)=u^{-2}$,
then, in evolutionary form, $\hat{\textbf{W}}_6=\left(U^2-TUU_T+\frac{U}{T^3}U_X\right)\frac{\partial}{\partial U}+F[U,V]\frac{\partial}{\partial V}$, where the differential function $F[U,V]$ must depend on $X$, $T$, $U$, $V$ and the partial derivatives of $U$ and $V$ with respect to $X$ and $T$. By applying $\hat{\textbf{W}}_6$ to the corresponding PDE system (\ref{SymB_eg nweq poten invereq}) which is invertibly related to the potential system (\ref{SymB_eg nweq potential}), one can show that $F[U,V]$ must be of the form $F(X,T,U,V,U_X,U_T)$. Applying $\hat{\textbf{W}}_6^{(\infty)}$ to the corresponding PDE system (\ref{SymB_eg nweq poten invereq}) and making appropriate substitutions, one can prove that the resulting determining equation system is inconsistent. Hence  $\textbf{W}_6$ yields a nonlocal symmetry of the potential system (\ref{SymB_eg nweq potential}) with $c(u)=u^{-2}$.

By similar reasoning, it turns out that $\textbf{W}_7$ also yields a nonlocal symmetry of the potential system (\ref{SymB_eg nweq potential}) with $c(u)=u^{-\frac{2}{3}}$. \hfill$\square$\medskip

When $c(u)$ is arbitrary, in $(x,t,u,v)$ coordinates, $\textbf{W}_3=(x+\int^uc^2(\xi)d\xi)\frac{\partial}{\partial x}+(t+v)\frac{\partial}{\partial t}$.
It is straightforward to show that $\textbf{W}_3$ is a point symmetry of a  potential system (\ref{SymB_eg nweq potential}) for arbitrary $c(u)$, whose infinitesimal component for the variable $t$ has an essential dependence on $v$. By projection, $\textbf{W}_3$ yields a nonlocal symmetry of a nonlinear wave equation (\ref{SymB_eg nweq}) for arbitrary $c(u)$.

When $c(u)=u^{-2}$, the infinitesimal components for the variables $(x,u)$ of the symmetry $\textbf{W}_6$ depend on the variable $v$. By Remark \ref{Rmk_symb nw V change}, $\textbf{W}_6$ yields a nonlocal symmetry of the corresponding nonlinear wave equation (\ref{SymB_eg nweq}).

Consider the case when $c(u)=u^{-\frac{2}{3}}$. If the symmetry $\textbf{W}_7$  yields a local symmetry $\tilde{\textbf{W}}_7$ of the corresponding nonlinear wave equation (\ref{SymB_eg nweq}), then $\tilde{\textbf{W}}_7=\check{\textbf{W}}_7+f[u]\frac{\partial}{\partial t}$, where the differential function $f[u]$ depends on $x$, $t$, $u$  and the partial derivatives of $u$ with respect to $x$ and $t$. Since $\rho^{-1}(\frac{\partial}{\partial V})=\frac{\partial}{\partial v}-\frac{\partial}{\partial t}$, when tracing back to the corresponding potential system (\ref{SymB_eg nweq potential}),
the infinitesimal component for the variable $v$ must be equal to $-f[u]$. Thus $\textbf{W}_7$ would also yield a local symmetry of the corresponding potential system (\ref{SymB_eg nweq potential}), which is a contradiction since  $\textbf{W}_7$  yields a nonlocal symmetry of the corresponding potential system (\ref{SymB_eg nweq potential}). Hence   $\textbf{W}_7$  yields a nonlocal symmetry of the nonlinear wave equation (\ref{SymB_eg nweq}) with $c(u)=u^{-\frac{2}{3}}$.

\begin{remark}\label{Rmk_symb nw V change2}\rm
 One can show that the symmetries $\textbf{W}_4$ and $\textbf{W}_5$  respectively yield point symmetries $\tilde{\textbf{W}}_4=\textbf{W}_4+(\mu+1)V\frac{\partial}{\partial V}$ and $\tilde{\textbf{W}}_5=\textbf{W}_5+V\frac{\partial}{\partial V}$ of the corresponding PDE systems (\ref{SymB_eg nweq poten invereq}).  In terms of $(x,t,u,v)$ coordinates, $\tilde{\textbf{W}}_4=u\frac{\partial}{\partial u}+(2\mu+1)x\frac{\partial}{\partial x}+(\mu+1)t\frac{\partial}{\partial t}+(\mu+1)v\frac{\partial}{\partial v}$
  and $\tilde{\textbf{W}}_5=\frac{\partial}{\partial
u}+2x\frac{\partial}{\partial x}+t\frac{\partial}{\partial t}+v\frac{\partial}{\partial v}$.
  Hence, by projection, $\textbf{W}_4$ and $\textbf{W}_5$ yield point symmetries of the nonlinear wave equations (\ref{SymB_eg nweq}) with $c(u)=u^{\mu}$ and $c(u)=e^u$ respectively.
\end{remark}
\begin{remark}\label{Rmk_symb nw V change3}\rm
\textit{Comparing the symmetries listed in} \cite{B2007}\textit{,} \textit{one sees that the symmetries} $\textbf{W}_6$ \textit{and} $\textbf{W}_7$ \textit{yield previously unknown nonlocal symmetries of the nonlinear wave equations (\ref{SymB_eg nweq}) with $c(u)=u^{-2}$ and $c(u)=u^{-\frac{2}{3}}$ respectively.}
\end{remark}

\section{Conclusion and open problems}
In this paper, we presented a new systematic symmetry-based procedure to construct
nonlocally related PDE systems (inverse potential systems) for a given PDE
system. The starting point for this method is any
point symmetry of a given PDE system.
Our new symmetry-based method yields previously unknown nonlocally related PDE systems for
nonlinear reaction-diffusion equations, nonlinear diffusion equations and nonlinear wave
equations as well as nonlocal symmetries for nonlinear diffusion and nonlinear wave equations.  Most importantly, through the symmetry-based method, we have obtained previously
unknown nonlocal symmetries for nonlinear wave equations with  $c(u)=u^{-2}$ or $c(u)=u^{-\frac{2}{3}}$.


  Potential systems are under-determined for a given PDE
system with more than two independent variables. It is known that
point symmetries of such potential systems cannot  yield nonlocal
symmetries of the given PDE system without additional gauge
constraints relating potential variables and their derivatives \cite{Ba2}.
In the case of three or more independent variables, the inverse potential systems generated by the symmetry-based method presented in this paper
involve natural gauge constraints due to their construction from curl-type conservation laws. Are there examples of such inverse potential systems, especially for given nonlinear systems of physical interest,  that yield nonlocal symmetries?



\section*{Acknowledgement}
We thank a referee for many valuable suggestions that improved this paper.


\begin{thebibliography}{99}
\bibitem{Olver}  P. J. Olver, {\it Applications of Lie Groups to Differential Equations}, GTM No. 107 (Springer, New York,  1986).
\bibitem{Bk1} G. W. Bluman and  S. Kumei, {\it Symmetries and Differential Equations}, Appl. Math. Sci. No. 81 (Springer, New York, 1989).
\bibitem{Cant} B. J. Cantwell, \textit{Introduction to Symmetry Analysis}, Cambridge Texts in Applied Mathematics (Cambridge University Press, Cambridge, UK, 2002).
\bibitem{Bca1} G. W. Bluman, A. F. Cheviakov and S. C. Anco, \textit{Applications of Symmetry Methods to Partial Differential Equations},
Appl. Math. Sci. No. 168 (Springer, New York, 2010).
\bibitem{Bluman3} G. W. Bluman, S. Kumei and G. J. Reid, J. Math. Phys. \textbf{29}, 806 (1988).
\bibitem{Akh_Ibra1987} I. S. Akhatov, R. K. Gazizov and N. H. Ibragimov, Mathematical Modeling \textbf{280}, 22 (1987) (in Russian).
\bibitem{Akh_Ibra1991} I. S. Akhatov, R. K. Gazizov and N. H. Ibragimov, J. Sov. Math. \textbf{55}, 1401 (1991).
\bibitem{Tsyfraa2005} I. Tsyfra, A. Napoli,  A. Messina and V. Tretynyk,  J. Math. Anal. Appl. \textbf{307}, 724 (2005).
\bibitem{Bluman_Cheviakov_Ivanova_JMP2006} G. W. Bluman, A. F. Cheviakov and N. M. Ivanova, J. Math. Phys. \textbf{47}, 113505 (2006).
\bibitem{B2007} G. W. Bluman and A. F. Cheviakov, J. Math. Anal. Appl. \textbf{333}, 93 (2007).
\bibitem{Cheviakov_JMP2008} A. F. Cheviakov, J. Math. Phys. \textbf{49}, 083502 (2008).
\bibitem{Gem} A. F. Cheviakov, Comput. Phys. Commun. \textbf{172}, 48 (2007).
\bibitem{Zhdanov2006} R. Zhdanov and V. Lahno, SIGMA \textbf{1}, Paper 009 (2005).


\bibitem{cb2} A. F. Cheviakov and G. W. Bluman, J. Math. Phys. \textbf{51}, 103521 (2010).
\bibitem{cb3} A. F. Cheviakov and G. W. Bluman, J. Math. Phys. \textbf{51}, 103522 (2010).
\bibitem{Dorodnitsyn1982} V. A. Dorodnitsyn, Zh. Vychisl. Mat. i Mat. Fiz. \textbf{6}, 1393 (1982) (in Russian).

\bibitem{Popovych_Ivanova_JPA2005a} R. O. Popovych and N. M. Ivanova, J. Phys. A \textbf{38}, 3145 (2005).












\bibitem{Ovsiannikov1959} L. V. Ovsiannikov,  Dokl. Akad. Nauk USSR \textbf{125}, 492 (1959) (in Russian).
\bibitem{bk1979}  G. W. Bluman and S. Kumei, J. Math. Phys. \textbf{21}, 1019 (1980).
\bibitem{Ames_Lohner_Adams_JNM1981} W. F. Ames, R. J. Lohner and E. Adams, Internat. J. Nonlinear Mech. \textbf{16}, 439 (1981).
\bibitem{bk1987}  G. W. Bluman and S. Kumei, J. Math. Phys. \textbf{28}, 307 (1987).


\bibitem{Ba2} S. C. Anco and G. W. Bluman, J. Math. Phys. \textbf{38}, 3508 (1997).


























\end{thebibliography}
\end{document}